\documentclass[fleqn,usenatbib]{mnras}

\usepackage{newtxtext,newtxmath}
\usepackage{mathtools}

\usepackage[T1]{fontenc}

\usepackage{graphicx}
\usepackage{amsmath}
\usepackage{soul}
\usepackage{xspace}
\usepackage{ulem}                   
\makeatletter 
  
  \patchcmd{\NAT@citex}
    {\@citea\NAT@hyper@{
      \NAT@nmfmt{\NAT@nm}
      \hyper@natlinkbreak{\NAT@aysep\NAT@spacechar}{\@citeb\@extra@b@citeb}
      \NAT@date}}
    {\@citea\NAT@nmfmt{\NAT@nm}
    \NAT@aysep\NAT@spacechar\NAT@hyper@{\NAT@date}}{}{}

  \patchcmd{\NAT@citex}
    {\@citea\NAT@hyper@{
      \NAT@nmfmt{\NAT@nm}
      \hyper@natlinkbreak{\NAT@spacechar\NAT@@open\if*#1*\else#1\NAT@spacechar\fi}
        {\@citeb\@extra@b@citeb}
      \NAT@date}}
    {\@citea\NAT@nmfmt{\NAT@nm}
    \NAT@spacechar\NAT@@open\if*#1*\else#1\NAT@spacechar\fi\NAT@hyper@{\NAT@date}}
    {}{}
\makeatother

\newcommand{\LCDM}{$\Lambda$CDM\xspace}

\newcommand{\hMsol}{h^{-1}\,{\rm M_\odot}}
\newcommand{\Msun}{{\rm M_\odot}}

\newcommand{\hMpc}{h^{-1}\,{\rm Mpc}}

\newcommand{\ie}{{i.e.~}}
\newcommand{\eg}{{e.g.~}}

\newcommand{\HI}{\ion{H}{I}\xspace}

\newcommand{\HeI}{\ion{He}{I}\xspace}
\newcommand{\HeII}{\ion{He}{II}\xspace}

\newcommand{\CIV}{\ion{C}{IV}\xspace}

\newcommand{\Lya}{Ly$\alpha$\xspace}
\newcommand{\taueff}{\tau_\mathrm{eff}}
\newcommand{\mfp}{\lambda_\mathrm{mfp}}
\newcommand{\avgf}{\langle F \rangle}
\newcommand{\meanf}{\bar{F}}
\newcommand{\fesc}{f_{\rm esc}}
\newcommand{\citenp}[1]{\citeauthor{#1} \citeyear{#1}}

\newcommand{\arepo}{{\sc arepo}\xspace}

\newcommand{\cloudy}{{\sc CLOUDY}\xspace}
\newcommand{\thesan}         {\textsc{thesan}\xspace}
\newcommand{\thesanone}      {\textsc{thesan-1}\xspace}
\newcommand{\thesantwo}      {\textsc{thesan-2}\xspace}
\newcommand{\thesanwc}       {\textsc{thesan-wc-2}\xspace}
\newcommand{\thesanlow}      {\textsc{thesan-low-2}\xspace}
\newcommand{\thesanhigh}     {\textsc{thesan-high-2}\xspace}

\newcommand{\thesansdao}     {\textsc{thesan-sdao-2}\xspace}
\newcommand{\thesantng}      {\textsc{thesan-tng-2}\xspace}

\newcommand{\thesansmall}    {\textsc{thesan-small}\xspace}
\newcommand{\thesansmalltng} {\textsc{thesan-small-tng}\xspace}
\newcommand{\dd}{\mathrm{d}}
\newcommand{\paperI}{Paper~\textsc{I}\xspace}
\newcommand{\paperII}{Paper~\textsc{II}\xspace}
\newcommand{\jwst}{\texttt{JWST}\xspace}
\newcommand{\alma}{\texttt{ALMA}\xspace}

\newcommand{\ska}{\texttt{SKA}\xspace}
\newcommand{\hera}{\texttt{HERA}\xspace}
\newcommand{\lofar}{\texttt{LOFAR}\xspace}
\newcommand{\ccatp}{\texttt{CCAT-p}\xspace}
\newcommand{\spherex}{\texttt{SPHEREx}\xspace}
\newcommand{\mwa}{\texttt{MWA}\xspace}

\newcommand{\highz}{\mbox{high-$z$}\xspace}

\graphicspath{{./}{figures/}}

\title[The high-redshift IGM in {\sc thesan}]{The \thesan\ project: properties of the intergalactic medium and its connection to reionization-era galaxies}

\author[E.~Garaldi et al.]{
E.~Garaldi,$^{1}$\thanks{E-mail: \href{mailto:egaraldi@mpa-garching.mpg.de}{egaraldi@mpa-garching.mpg.de}}
R.~Kannan,$^{2}$\thanks{E-mail: \href{mailto:rahul.kannan@cfa.harvard.edu}{rahul.kannan@cfa.harvard.edu}}
A.~Smith,$^{3}$\thanks{E-mail: \href{mailto:arsmith@mit.edu}{arsmith@mit.edu}; NHFP Einstein Fellow.}
V.~Springel,$^{1}$
R.~Pakmor,$^{1}$
M.~Vogelsberger$^{3}$
and L.~Hernquist$^{2}$
\\
$^{1}$Max-Planck Institute for Astrophysics, Karl-Schwarzschild-Str.~1, D-85741 Garching, Germany\\
$^{2}$Center for Astrophysics $\vert$ Harvard $\&$ Smithsonian, 60 Garden Street, Cambridge, MA 02138, USA\\
$^{3}$Department of Physics, Massachusetts Institute of Technology, Cambridge, MA 02139, USA\\
}

\date{Accepted XXX. Received YYY; in original form ZZZ}

\pubyear{2022}

\begin{document}
\label{firstpage}
\pagerange{\pageref{firstpage}--\pageref{lastpage}}
\maketitle

\begin{abstract}  
The high-redshift intergalactic medium (IGM) and the primeval galaxy population are rapidly becoming the new frontier of extra-galactic astronomy. 
We investigate the IGM properties and their connection to galaxies at $z\geq5.5$ under different assumptions for the ionizing photon escape and the nature of dark matter, employing our novel \thesan radiation-hydrodynamical simulation suite, designed to provide a comprehensive picture of the emergence of galaxies in a full reionization context. 
Our simulations have realistic `late' reionization histories, match available constraints on global IGM properties and reproduce the recently-observed rapid evolution of the mean free path of ionizing photons. 
We additionally examine \highz Lyman-$\alpha$ transmission. 
The optical depth evolution is consistent with data, and its distribution suggests an even-later reionization than simulated, although with a strong sensitivity to the source model. 
We show that the effects of these two unknowns can be disentangled by characterising the spectral shape and separation of Lyman-$\alpha$ transmission regions, opening up the possibility to observationally constrain both. 
For the first time in simulations, \thesan reproduces the modulation of the Lyman-$\alpha$ flux as a function of galaxy distance, demonstrating the power of coupling a realistic galaxy formation model with proper radiation-hydrodynamics. 
We find this feature to be extremely sensitive on the timing of reionization, while being relatively insensitive to the source model. 
Overall, \thesan produces a realistic IGM and galaxy population, providing a robust framework for future analysis of the \highz Universe.
\end{abstract}


\section{Introduction}
\label{sec:intro}

Within a billion years of the Big Bang, the gas content of the Universe changed its ionization state twice: it became neutral at the recombination epoch as a consequence of the cooling provided by the expansion of the Universe (producing the Cosmic Microwave Background, or CMB), and it was later ionized again (most likely) by the first generations of galaxies during a period known as the Epoch of Reionization \citep[EoR, \eg][]{Shapiro&Giroux1987,Loeb&Barkana2001,Haiman2016}.

Reionization is the least understood period of time in the history of the Universe, because of extreme observational difficulties in probing this remote epoch and formidable challenges to its theoretical investigation. However, understanding its unfolding is essential, as it constitutes the evolutionary link between the high-redshift Universe and the cosmic structures today. As such, it plays an important role in the origin of galaxies and provides a test-bed for evaluating our knowledge --~mostly gained from local observations~-- of the relevant physical processes in galaxy formation.

From decades of investigations, a relatively-coherent picture of the EoR has been built. The modeling of anisotropies in the CMB \citep{Planck2018cosmo}, the evolution of the Lyman-$\alpha$ (hereafter \Lya) forest opacity \citep{Becker+2001,Fan+2006,Becker+2015}, and of the number density of detected \Lya-emitting systems \citep{Ota+2010,Pentericci+2011,Mason+2018}, the damping wings of high-redshift quasars \citep[\eg][]{Schroeder+2013, Greig+2017, Davies+2018}, and the statistics of dark pixels in quasar absorption spectra \citep[\eg][]{McGreer+2011} have allowed us to constrain the EoR to the redshift range $5.5 \lesssim z \lesssim 10$ and its tail-end (where individual ionized regions coalesce) to occur at $5 \lesssim z \lesssim 6$. 
It is now mostly accepted that this process was mainly powered by a swarm of small (\ie with stellar mass $M_\mathrm{star} \lesssim 10^8 \, \Msun$), star-forming galaxies with relatively high escape fractions of ionizing photons \citep{madau99,gnedin00,Haardt&Madau12}, although in recent times some evidence against this picture has been presented \citep{Naidu+2020}. Active galactic nuclei \citep[AGN, ][]{Haardt&Madau1996,Kulkarni+2018} are thought to have played only a minor role at redshifts $z\gtrsim5$ (but see \citenp{madau15}; \citenp{Garaldi+2019}). 

Observationally confirming this picture, however, requires understanding the evolution of the intergalactic medium (IGM) between galaxies as well as the formation of the first structures. This task is arduous. 
For instance, to date only $\mathcal{O}(1000)$ galaxy candidates have been discovered at $6 \lesssim z \lesssim 8$ \citep[and much fewer at higher redshift, see \eg][]{Bouwens+2015,Livermore+2017,Atek+2018}, many of which do not yet have spectroscopic characterisations. 
Nevertheless, the formation of primeval galaxies above redshift $z \gtrsim 6$ is rapidly becoming the new frontier in the study of both cosmic reionization and galaxy formation, as evidenced by recent observational campaigns that have started to constrain the \textit{population} of galaxies at the end of the EoR, \eg REBELS \citep{rebels} and ALPINE \citep{alpine}. 
These are (or will soon be) flanked by many instruments providing us with a multi-faceted picture of the emergence of primeval galaxies in the context of the EoR. 
Among them, the Atacama Large Millimiter Array (\alma), the \textit{James Webb Spaces Telescope} (\jwst), the Cerro Chajnantor Atacama Telescope-prime(\ccatp), and the Spectro-Photometer for the history of the Universe, Epoch of Reionization and Ices Explorer (\spherex) focus on the properties of \highz galaxies, while the Low-Frequency Array (\lofar), the Murchison Widefield Array (\mwa), the Hydrogen Epoch of Reionization Array (\hera), and the Square Kilometer Array (\ska) investigate instead the distribution of neutral hydrogen before and during the EoR, either statistically or tomographically.

These new-generation instruments are designed to assist the traditional means of investigations of the EoR, which mainly relied on the \Lya transition in neutral hydrogen atoms, the most prominent and ubiquitous absorption line in quasar spectra. 
However, because of its large absorption cross section, even a modest hydrogen neutral fraction of $x_\mathrm{\HI} \sim 10^{-4}$ completely suppresses the incoming flux. This has limited the reach of \Lya absorption studies to the post-reionization Universe until recent years, when technological advancements (in particular better spectral resolution and sensitivity) coupled with the discovery of QSOs at higher and higher redshift have allowed us to probe the end phases of the EoR \citep[\eg][]{Barnett+2017,Kakiichi+2018,Meyer+2019,Bosman+2018,Eilers+2018,Yang+2020}.

These advancements provided new fuel for refined theoretical investigations. For instance, \citet{Kulkarni2019} and \citet{Keating+2020} combined these observations with post-processing radiation transport (RT) simulations to provide evidence in support of a completion of the reionization process as late as $z \sim 5.2$ (now known as a `late' reionization model), which appears necessary to match the distribution of optical depth at $z \lesssim 6$. 
\citet{Gnedin+2017}, \citet{Garaldi+2019croc}, and \citet{Gaikwad+2020} all characterised the properties of the \Lya transmission regions at $z \lesssim 6.5$ in order to establish the fidelity of current state-of-the-art simulations and connect the geometrical properties of such transmission regions to the underlying physical conditions of the IGM between galaxies.
Notably, \citet{Garaldi+2019croc} demonstrated that even state-of-the-art radiation-hydrodynamical simulations are unable to match the observed connection between \Lya transmission in QSO spectra and the galaxy population at $z\sim 5.5$. 

The aforementioned work represents an example of how the connection between cosmic reionization and the galaxy population is one of the most challenging and promising grounds for the investigation of the \highz Universe. In fact, the convergence of the fields of cosmic reionization and galaxy formation poses a fierce new challenge; \ie to simultaneously understand the properties of the intergalactic medium (IGM) on large scales and of high-redshift galaxies on small scales, as well as --~crucially~-- their connection in a unique and consistent picture. 

In an attempt to improve our understanding of the EoR, a number of different numerical approaches have been developed. However, numerical studies of the EoR are among the most challenging in the landscape of numerical astrophysics. 
The reason lies in the requirement to simultaneously: (i) cover large volumes, in order to provide a statistically-representative picture of the reionization process; (ii) resolve (sub-)galactic scales, 
where processes responsible for the production of ionizing photons occur, where the self-shielded gas acting as photons sink resides, and where galactic observables are produced; and (iii) include an accurate treatment of RT. 
The gargantuan computational cost of simulations with these characteristics has forced most of the efforts to date to employ simplifying approximations. 
In fact, while 
fully-coupled radiation-hydrodynamics simulations have become possible 
in recent years, they are often limited to small volumes or individual galaxies \citep[\eg][]{Aurora, SPHINX, Pallottini2017, Katz2019, Renaissance, Obelisk}, or coarse resolutions \citep[\eg][]{CoDa, CoDaII}, although notable exceptions exist \citep{CROC}. 
A common issue of these simulations is the fact that they are too expensive to be run to $z=0$, and therefore their galaxy formation models are only calibrated and tested against $z\gtrsim5$ (scarce and of limited quality) observations. This renders the fidelity of the simulated physics uncertain. In fact, when some of these models have been evolved to lower redshift, they produced galaxy properties incompatible with observed ones \citep[\eg][]{Tacchella2018, Trebitsch+2020, Mitchell2021}.

In order to achieve a comprehensive and statistically-significant view of the \highz Universe and bypass the aforementioned problems, we have developed the \thesan simulation suite, which we introduce in this and two companion papers. In particular, \citet[][hereafter \paperI]{Thesan_intro} presents an overview of the simulations, while \citet[][hereafter \paperII]{Thesan_Lya} describes results concerning the \Lya transmission from individual haloes. This paper focuses on the properties of the diffuse IGM and its \Lya forest in the \thesan suite, including a thorough comparison with available observations, shedding new light on the reionization process and the physics of high-redshift Universe. 
\thesan provides a unique combination of large simulated volumes, high numerical resolution, a wealth of physical processes simulated in a self-consistent manner (including RT and cosmic dust), exploration of high-redshift physics (including escape fraction and dark matter models), and a realistic galaxy formation model at \textit{all} redshifts. 
We intend that this suite will provide a firm ground upon which future investigations can be built. For this reason, we will make the data public in the near future and welcome proposals for collaborative work.

The paper is organised as follows. In Section~\ref{sec:methods} we present a brief overview of the simulation suite and details of the forward modelling performed in this study. In Section~\ref{sec:IGM} we describe results concerning the global properties of the IGM. In Section~\ref{sec:Lya} we focus on the \highz \Lya transition. In Section~\ref{sec:igm-gal-connetion} we investigate the connection between the \highz IGM and coeval galaxy population. Finally, we provide a summary and concluding remarks in Section~\ref{sec:conclusions}.

\begin{table*}
	\centering
	\caption{A summary of the main properties of the \thesan simulations employed in this paper. From left to right the columns report the name of the simulation, box size, initial particle number, mass of the dark matter particles and gas cells, the (minimum) softening length of (gas) star and dark matter particles, minimum cell size at $z=5.5$, the final redshift, the escape fraction of ionizing photons from the birth cloud (if applicable) and a short description of the simulation. A complete version of this Table can be found in \paperI.}
	\label{table:simulations}
	\addtolength{\tabcolsep}{-0.5pt}
	\begin{tabular}{lccccccccl} 
		\hline
		Name & ${L}_\mathrm{box}$ & $N_\mathrm{particles}$ & ${m}_\mathrm{DM}$ & $m_\mathrm{gas}$ & $\epsilon$ & $r^\mathrm{min}_\mathrm{cell}$& $z_\mathrm{end}$ & $f_\mathrm{esc}$ & Description\\  
		& [cMpc] & & [$\mathrm{M}_\odot$] & [$\mathrm{M}_\odot$] & [ckpc] & [pc] & & &\\
		\hline
		\thesanone & $95.5$  & $2 \times 2100^3$ & $3.12 \times 10^6$ & $5.82 \times 10^5$ & $2.2$ & $\sim 10$ & $5.5$ & $0.37$ & fiducial model (RMHD + TNG + dust) \\
		\\
		\thesantwo & $95.5$  & $2 \times 1050^3$ & $2.49 \times 10^7$ & $4.66 \times 10^6$ & $4.1$ & $\sim 35$ & $5.5$ & $0.37$ & fiducial model\\
		\thesanwc  & $95.5$  & $2 \times 1050^3$ & $2.49 \times 10^7$ & $4.66 \times 10^6$ & $4.1$ & $\sim 35$ & $5.5$ & $0.43$ & fiducial model + weak convergence of $x_\mathrm{HI} (z)$\\
		
		\thesanhigh & $95.5$  & $2 \times 1050^3$ & $2.49 \times 10^7$ & $4.66 \times 10^6$ & $4.1$ & $\sim 35$ & $5.5$ & $0.8$ & fiducial model + $f_{\mathrm{esc}} = 0$ if $M_\mathrm{halo} \geq 10^{10} \hMsol$\\
		\thesanlow & $95.5$  & $2 \times 1050^3$ & $2.49 \times 10^7$ & $4.66 \times 10^6$ & $4.1$ & $\sim 35$ & $5.5$ & $0.95$ & fiducial model + $f_{\mathrm{esc}} = 0$ if $M_\mathrm{halo} < 10^{10} \hMsol$\\
	    
		\thesansdao & $95.5$  & $2 \times 1050^3$ & $2.49 \times 10^7$ & $4.66 \times 10^6$ & $4.1$ & $\sim 35$ &  $5.5$ & 0.55 & fiducial model + sDAO dark matter model\\
		\thesantng & $95.5$  & $2 \times 1050^3$ & $2.49 \times 10^7$ & $4.66 \times 10^6$ & $4.1$ & $\sim 35$ &  $5.5$ & - & MHD + TNG model (original TNG)\\
		\hline
	\end{tabular}
	\addtolength{\tabcolsep}{0.5pt}
\end{table*}

\begin{figure*}
\includegraphics[width=\textwidth]{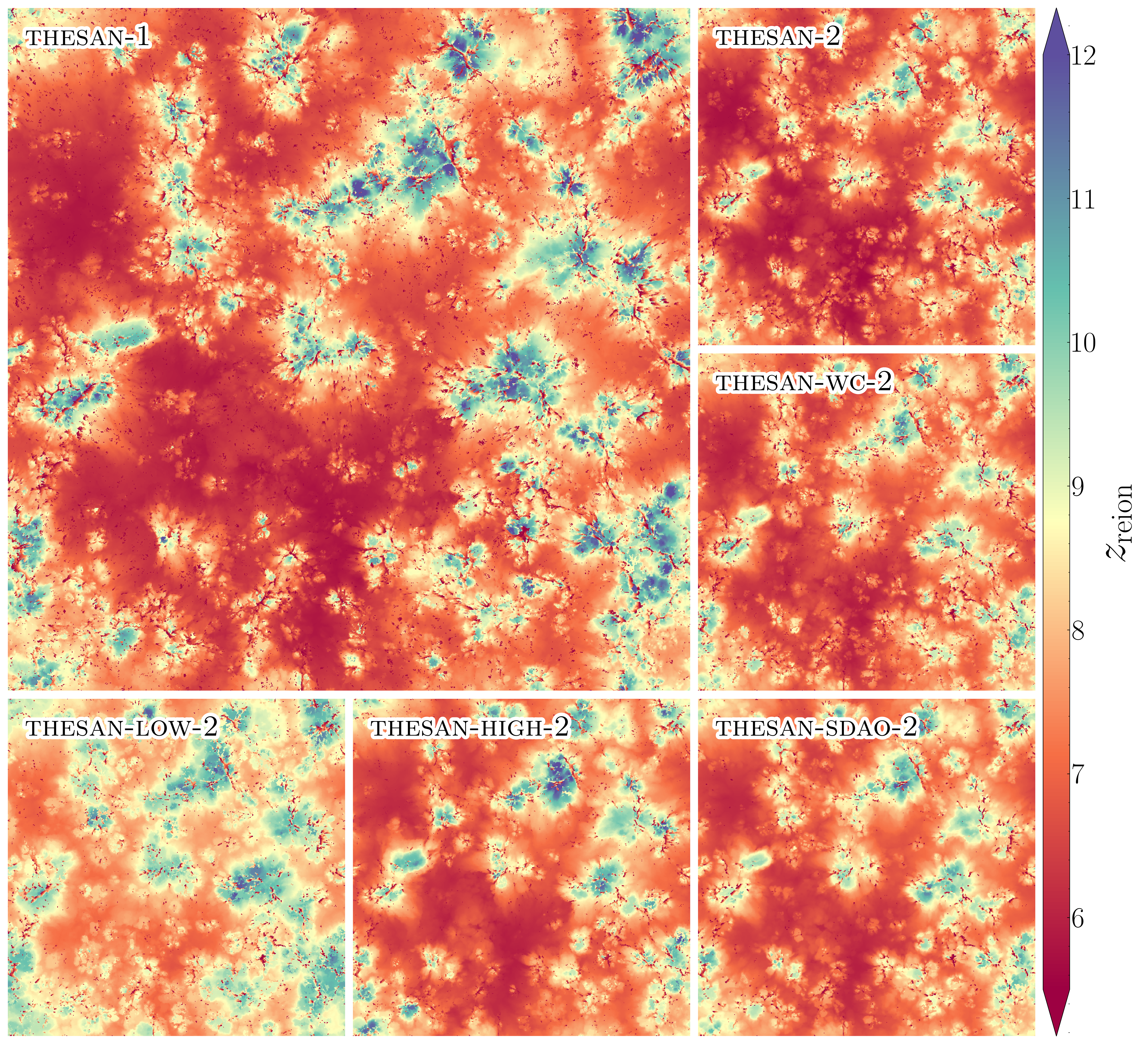}
\caption{Reionization redshift, defined as the latest time at which the ionized fraction of the resolution element crosses the threshold value $0.99$, for each resolution element in a slice through a selection of the \thesan simulations.  The different escape fraction and dark matter models produce visible differences in the reionization redshift distribution. Encouragingly, the fiducial and weak convergence runs show very few differences.}
\label{fig:z_reion}
\end{figure*}

\begin{figure*}
\centering
\includegraphics[width=\textwidth]{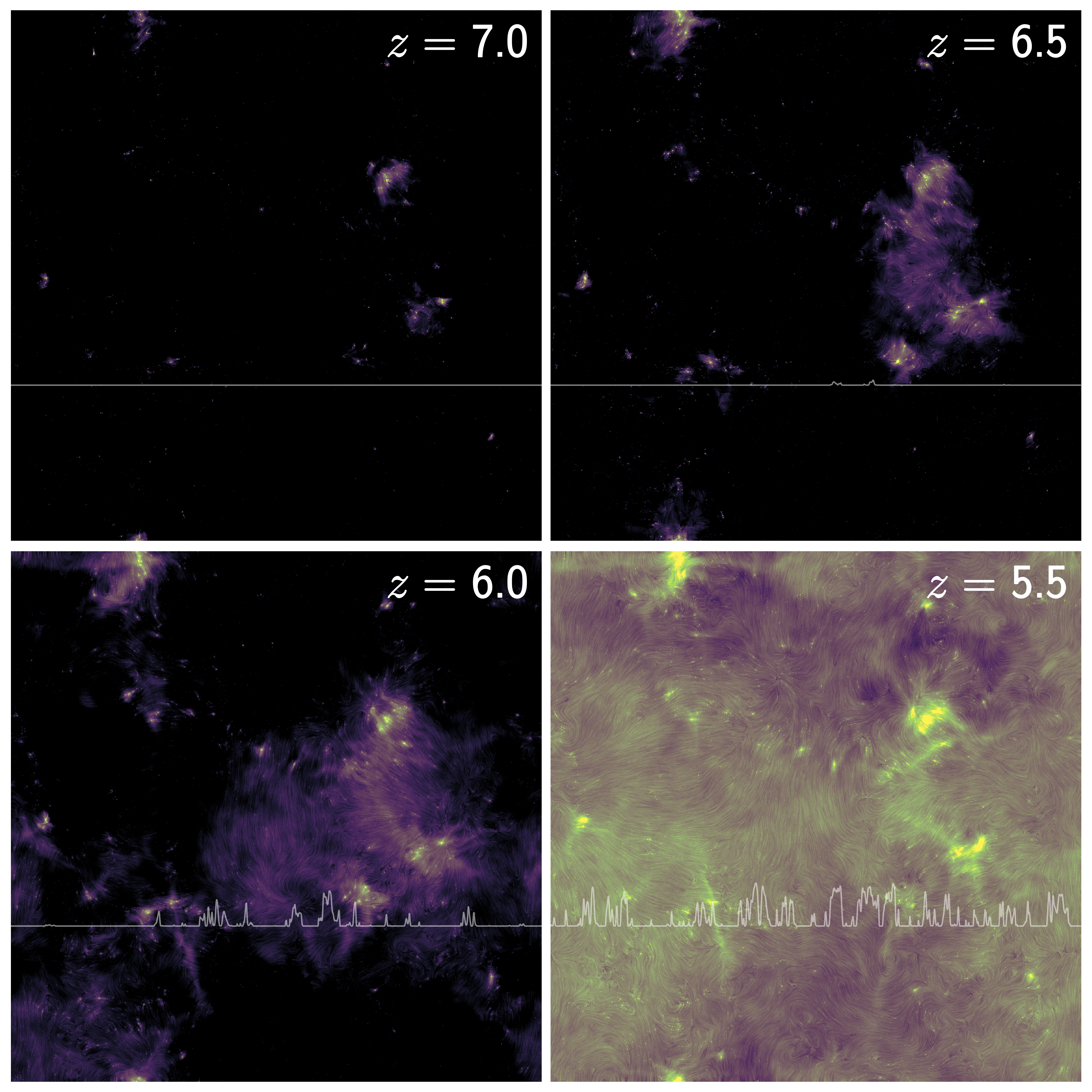}
\caption{Visualization of a slice (of thickness equal to $20$\% of the box size) through \thesanone. Panels show the line integral convolution of the photon density (providing the background texture) and the photon flux along the image plane (providing the vector field). In each panel, the redshift is indicated in the top right corner. Galaxies and galaxy groups act as sources of photons, which travel approximately linearly until they encounter radiation sinks. Notice that the color scale employed is identical in all panels. The change in the overall hue of the panels reflects the rapid evolution of the ionising photon field in the latest phases of reionization. Additionally, we show in each panel the \Lya transmitted flux in a single sightline traversing the slice, showing the rapid build-up of transmitted flux in ionised regions at $z \lesssim 6$.}
\label{fig:lic}
\end{figure*}

\section{Methods}
\label{sec:methods}

In this Section we summarise the main features of the \thesan simulations, which are more thoroughly described in \paperI. These represent the backbone of the analysis presented here, as they provide the physical state of the simulated universe. In order to fruitfully compare simulated and observed quantities, we post-process our simulations to produce synthetic observations that are as similar as possible to real data. We describe how this is achieved at the end of this Section.

\subsection{The \thesan simulations}
\label{sec:simulations}

In this work we employ the \thesan simulation suite of radiation-magneto-hydrodynamical simulations run with the moving-mesh hydrodynamics code \arepo \citep{Arepo,Arepo-public}\footnote{Public code access and documentation available at \href{https://arepo-code.org}{\texttt{www.arepo-code.org}}.}. 
The equations of radiation-magneto-hydrodynamics are solved on a mesh corresponding to the Voronoi tessellation of mesh-generating points that approximately follow the gas flow. 
The computation of gravity relies on a hybrid Tree-PM approach, which separately computes long-range (via a particle mesh approach) and short-range \citep[through a hierarchical oct-tree, ][]{Barnes&Hut86} forces. 
In addition, we employ a hierarchical time integration, allowing an efficient management of the deep time bin hierarchy, and a randomization of the node centres at each domain decomposition \citep[described in the \textsc{Gadget4} paper, ][]{Gadget4} that suppresses force errors at the boundary of large nodes.

\subsubsection{Galaxy formation model}
The \thesan simulations were designed to simultaneously capture the assembly of primeval galaxies and their coupling to reionization, while maximising their physical fidelity. 
For this reason, we choose to employ a sub-resolution physics model that has been extensively investigated at low redshift, namely the IllustrisTNG model \citep{TNG_Nelson,TNG_Pillepich,TNG_Springel,TNG_Marinacci,TNG_Naiman,Nelson2019a,Pillepich2019}. (Details of the TNG model can be found in \citenp{Weinberger2017}, \citenp{Pillepich2018b}, \citenp{Nelson2019b}. 
The model is an improvement over the Illustris one, described in \citenp{Illustris} and \citenp{Illustris_nature}. We note here that this model employs a tabulated spatially-uniform UV background, that we replace with self-consistent RT as described in the next Section.) This choice ensures that, although the \thesan simulations are not evolved below $z=5.5$, their physical model and the galaxy population they produce can be trusted throughout the entire history of the Universe. 
We present supporting evidence for this claim in \paperI. In particular, we show that \thesan is able to produce a realistic stellar-to-halo-mass relation (figure 9 of \paperI), stellar mass function (figure 10 of \paperI), UV luminosity function (figure 11 of \paperI), star-formation-rate-density evolution (figure 12 of \paperI), and mass-metallicity relation (figure 14 of \paperI). 
In addition, \citet{Wu+2019} demonstrated in a very similar configuration that replacing the \citet{Faucher-Giguere+09} UV background (UVBG) with self-consistent RT does not significantly alter the main properties of galaxies, while having a large impact on the properties of the IGM. 
Finally, we check in Appendix~\ref{sec:tng_comparison} that this holds true in our simulations as well by employing the \thesantwo and \thesantng runs. All these results make us confident that the successes of the IllustrisTNG model will be retained in the \thesan model at all redshifts. 
Additionally, employing the IllustrisTNG model \textit{as is} (with the exception of the UVBG being replaced by the self-consistent RT) effectively limits the free parameters to a single number, the (unresolved) stellar escape fraction $f_\mathrm{esc}$ (which quantifies the sub-resolution absorption and is not present in the IllustrisTNG model as it lacks RT), which we calibrate against the reionization history of the Universe. In fact, all other physical parameters have been tested against low-redshift observables during the development of the IllustrisTNG simulations. 
We note here that, among other physical processes, the IllustrisTNG model and therefore the \thesan simulations as well include black holes, and  magnetic fields originating from a uniform, cosmological seed field of initial value $B_\mathrm{seed} = 10^{-14}$ G. The former are created placing a seed black hole of mass $M_\mathrm{seed} = 8 \times 10^5 \, \hMsol$ at the center of every FOF group that exceeds a mass threshold of $M_\mathrm{thr} = 5 \times 10^{10} \, \hMsol$. They then accrete material from neighbouring cells using a Bondi-Hoyle prescription \citep{Bondi&Hoyle1944, Bondi1952} and release energy into the surrounding gas via two feedback modes \citep[dubbed quasar and radio mode, see][]{Weinberger2017}. We refer the reader to \paperI for an overview of the black hole population in the \thesan simulations, and remark here that they globally contribute less than $1$\% of the hydrogen-ionizing photons.

\subsubsection{Self-consistent radiative transfer}
In order to self-consistently solve the RT equations, we employ the \textsc{arepo-rt} \citep{ArepoRT} extension of the \arepo code, which solves the first two moments of the RT equation, complemented with the M1 closure relation \citep{Levermore84}. 
Because of computational limitations, we decide to follow only the UV part of the radiation spectrum and discretise the photon energy into bins defined by the following energy thresholds: $[13.6, 24.6, 54.4,\infty)$\,eV, chosen to braket the ionization energies of hydrogen and helium. Each resolution element tracks, for each bin, the comoving photon number density and flux. 
In order to partially compensate the loss of resolution in the frequency sampling, which may introduce errors in the temperature and ionization state of the gas \citep[see \eg][]{Mirocha+2012}, we assume the radiation follows, within each bin, the spectrum of a $2$\,Myr old, quarter-solar metallicity stellar population with amplitude given by the photon number density of the cell itself. 
The values of the integral average over the bin of the \HI, \HeI and \HeII cross section ($\sigma_X$), energy injected per unit photon ($\mathcal{E}_X$), and photon energy ($e$) are reported in Table~1 of \paperI. 
The stellar spectra are computed as a function of mass, metallicity and age of each stellar particle using the Binary Population and Spectral Synthesis models v2.2.1 \citep[BPASS, ][]{BPASS2017, BPASS2018}, assuming a Chabrier IMF \citep{Chabrier2003}. Additionally, we include radiation from quasars by converting the black hole accretion rate computed by the code into a quasar luminosity (assuming a radiative efficiency of $\eta_\mathrm{rad} = 0.2$) and assuming the spectral shape from \citet{Lusso+2015} with unit escape fraction. 
Finally, we employ a reduced speed of light (RSLA) approximation, with an effective value of the speed of light $\Tilde{c} = 0.2 c$. Recently, there has been a debate in the literature on whether employing a RSLA produces appreciable effects on the reionization of the Universe \citep{Bauer+2015, Gnedin2016rsla, Deparis+19, Ocvirk+19}. We have carefully chosen the value of the (reduced) speed of light to prevent any sizeable effect on the IGM (as measured by the reionization front propagation speed and reionization history, see Appendix A of \paperI). We are therefore confident that the reionization properties in \thesan are not affected by this approximation. Additionally, \citet{Gnedin2016rsla} showed that the RSLA approximation does not impact  star-formation properties of galaxies for our employed value of $\Tilde{c}$, supporting our expectation that the galaxy properties are robust even when employing the RSLA. Overall, we believe that the use of a (mild) RSLA does not have appreciable effects on the results presented in this paper. 

The RT equations are coupled to a non-equilibrium solver that accurately computes the ionization state of hydrogen and helium, as well as the temperature change due to photo-ionization, metal cooling and Compton cooling. Finally, in order to provide a more faithful ground of comparison with observations of \highz dust, we have included in the \thesan simulation the dust model developed by \citet{McKinnon+16,McKinnon+17}. 

The inclusion of a realistic reionization history implies that galaxies in \thesan are subject to a spatially-varying UV radiation field. This is expected to affect galaxies by means of radiative feedback on the gas (\eg radiation pressure, photo-ionization and photo-heating) in a different way than the combination of spatially-uniform UVBG and approximate self-shielding corrections that is usually employed in cosmological hydrodynamical simulations (including in the IllustrisTNG model). 
As discussed, however, \thesan describes the ISM through an effective equation of state \citep[eEOS, originally developed in][]{Springel&Hernquist03}. 
In this model, the ISM gas is assumed to be composed of a volume-filling hot and diffuse component in pressure equilibrium with a cold and dense phase, where stars are formed. From the gas density, its temperature, star-formation rate and other properties can be inferred. 
The advantage of such a treatment is that the global properties of the ISM gas are captured without the need to resolve physical processes on small scales, making it possible to efficiently run accurate galaxy simulations of cosmological volumes \citep[see \eg the IllustrisTNG simulations, or the Auriga suite,][]{Auriga}. 
However, this simplicity comes at the cost of an ISM composition that is assumed a priori and involves a tight relation between gas density and temperature. 
This rigidity means that feedback processes can affect the gas in only two ways, namely they either: (i) heat the gas enough to prevent it from cooling back to the eEOS temperature within a single time step of the simulation, hence preventing star formation for the time step, or (ii) they affect the gas \textit{outside} the galaxy, altering its accretion onto it. The latter occurs through \eg photo-heating of the diffuse gas inside and outside the halo \citep[\eg][]{Okamoto+08}. 
Therefore, simulating the ISM using an eEOS model partially hides the effects of radiative feedback on the small-scale structure of the ISM. Nevertheless, this approach captures the indirect effects of photo-heating, as shown by \citet{Wu+2019} and discussed in the previous section as well as in Appendix~\ref{sec:tng_comparison}. Additionally, galactic properties are well captured, as shown by the large body of simulations employing this approach and reproducing many observations.

\subsubsection{Simulation design}
All \thesan simulations have a box size of $L_\mathrm{box} = 95.5 \, \mathrm{cMpc}$ and assume a \citet{Planck2015cosmo} cosmology. Our flagship simulation (\thesanone) contains $2\times2100^3$ particles, reaching dark matter and gas resolutions of $m_\mathrm{DM} = 3.12 \times 10^6 \, \Msun$ and  $m_\mathrm{gas} = 5.82 \times 10^5 \, \Msun$, respectively. This allows us to resolve atomic cooling haloes ($M_\mathrm{halo} \sim 10^8 \, \hMsol$). The forces are softened on a scale of $2.2 \, \mathrm{ckpc}$, while the smallest gas resolution elements reach sizes of approximately $10 \, \mathrm{pc}$. This run is accompanied by a number of medium-resolution runs, covering the same patch of the universe, but featuring slightly coarser resolution (\ie $8$ times lower mass resolution and $2$ times lower spatial resolution). 
While the full description of these runs can be found in \paperI, we report in Table~\ref{table:simulations} the main features of the subset of runs employed in this Paper. 
In particular, the \thesantwo run has the same physical model as the flagship one, while in \thesanwc we have adjusted the value of the stellar unresolved escape fraction to compensate the lower star-formation rate and keep the reionization history approximately identical to \thesanone. In \thesanlow and \thesanhigh we force galaxies only above and below $M_\mathrm{halo} = 10^{10} \, \Msun$ --~respectively~-- to contribute to the EoR. In \thesansdao, we replace the \LCDM dark matter model with one featuring strong dark acoustic oscillations \citep[\eg][]{Bohr2020}. Finally, \thesantng employs the original IllustrisTNG model, which does not include full radiative transfer.

\begin{figure*}
\includegraphics[width=\textwidth]{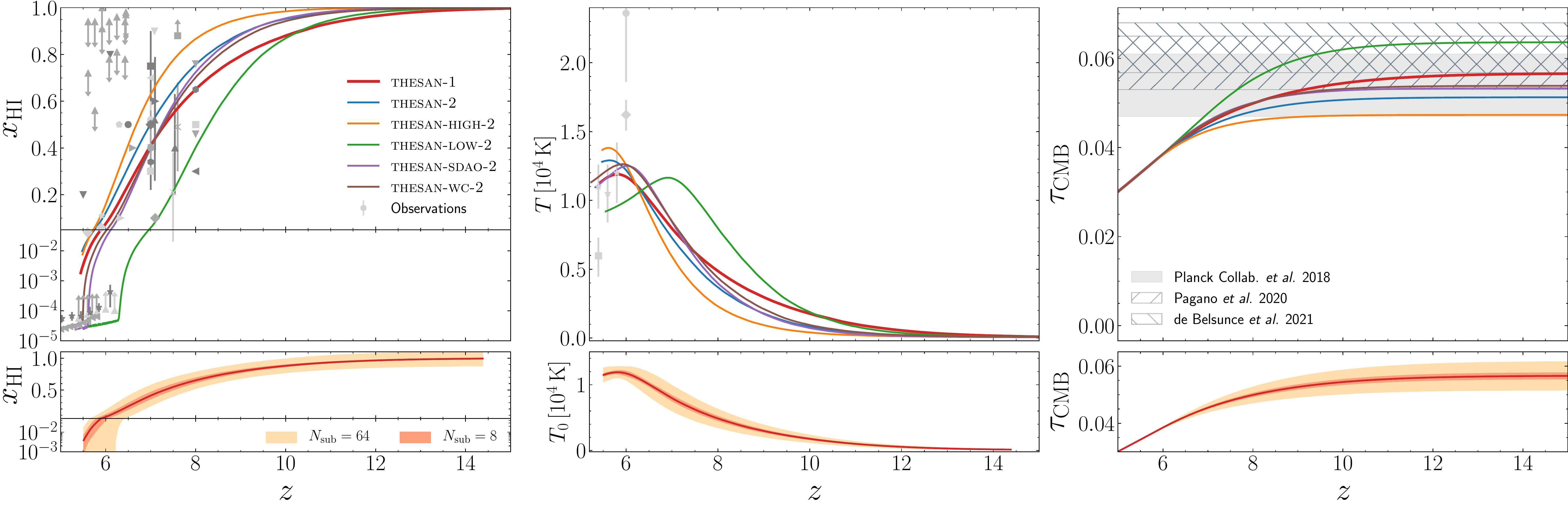}
\caption{Global properties of the IGM in the \thesan simulations, compared to a collection of observations from the literature. From left to right: volume-averaged hydrogen neutral fraction evolution, volume-averaged temperature at mean density evolution, and CMB Thomson optical depth as a function of maximum integration redshift. The bottom panels show the variance around the \thesanone curve, computed by subdividing the original box into $8$ (inner shaded region) and $64$ (outer shaded region) sub-boxes. Overall, most \thesan simulations agree well with available data. Constraints on the neutral fraction are derived from the evolution of \Lya emitters and the number density of Lyman break galaxies (\citenp{Ota+2008} -- light grey circles, \citenp{Ono+2012} -- dark grey squares, \citenp{Jensen+2013} -- dark grey circles, \citenp{Pentericci+2014} -- light grey hexagons, \citenp{Tilvi+2014} -- dark grey left-pointing triangles, \citenp{Choudhury+2015} -- light grey squares, \citenp{Mesinger+2015} -- grey circles), from the dark pixel statistics (\citenp{McGreer+2011} -- dark grey down-pointing triangles, \citenp{McGreer+2015} -- light grey diamonds, \citenp{Lu+2020} -- grey up-pointing triangles), from GRB afterglow (\citenp{Totani+2006} -- light grey pentagons, \citenp{Chornock+2013} -- light grey stars, from the Doppler widths of \Lya absorption lines in the quasar near zones (\citenp{Bolton+2012} -- light grey down-pointing triangles), from the quasar damping wing (\citenp{Mortlock+2011} -- grey diamonds, \citenp{Schroeder+2013} -- light grey right-pointing triangles, \citenp{Greig+2017} -- dark grey right-pointing triangles, \citenp{Greig+2019} -- light grey crosses, \citenp{Wang+2020} -- light grey left-pointing triangles), from CMB modeling (\citenp{Robertson+2013} -- dark grey diamonds), from the Gunn-Peterson optical depth (\citenp{Fan+2006} -- dark grey stars, \citenp{Davies+2018} -- dark grey up-pointing triangles, \citenp{Yang+2020} -- light grey up-pointing triangles, \citenp{XQR30} -- grey left-pointing triangles), from the angular correlation function of \Lya emitters (\citenp{Sobacchi+Mesinger2015} -- dark grey crosses), from the rest-frame UV continuum of galaxies (\citenp{Schenker+2014} -- dark grey hexagons), from the detection of \Lya emission in Lyman break galaxies (\citenp{Mason+2018} -- grey stars, \citenp{Mason+2019} -- grey down-pointing triangles, \citenp{Hoag+2019} -- grey squares, \citenp{Jung+20} -- dark-grey crosses), from a combination of \Lya luminosity function, clustering and line profile (\citenp{Ouchi+2010} -- grey right-pointing triangles) and from the \Lya visibility (\citenp{Dijkstra+2011} -- dark grey pentagons). Constraints on $T_0$ are derived from the Doppler widths of \Lya absorption lines in the quasar near zones (\citenp{Bolton+2010} -- light-grey circles, \citenp{Bolton+2012} -- light grey diamonds, from the \Lya forest power spectrum (\citenp{Walther+2019} -- light-grey squares), and from the Doppler widths of \Lya transmission regions (\citenp{Gaikwad+2020} -- light grey stars).}
\label{fig:igm_prop}
\end{figure*}

\subsection{Synthetic spectra}
\label{sec:spectra}
The simulation outputs were post-processed to produce realistic synthetic spectra. We have developed a low memory method for exact ray-tracing through the native Voronoi unstructured mesh data as described in \paperII. We employ the Cosmic Ly$\alpha$ Transfer code \citep[\textsc{colt};][]{Smith+15,Smith+19} to extract the gas physical properties along 150 unique lines of sight (LOS) per snapshot. Each LOS has a length of $100 \, \mathrm{cMpc}$, originates from a randomly-chosen position in the simulation box, and propagates along a random direction employing periodic boundary conditions. These LOS are then processed to obtain the normalized transmitted flux for different transitions. We note here that 
we employ the full Voigt-Hjerting line profile (\citenp{Hjerting1938}, using the approximation presented in \citenp{Harris1948} and \citenp{Tepper-Garcia2006}) including the effects of gas temperature and peculiar velocities and produce spectra with spectral resolution of $\Delta \nu = 1 \, \mathrm{km\,s}^{-1}$. 
When comparing to a specific observational dataset, we additionally forward model these `ideal' spectra by including noise \citep[we make the assumption that the noise level is independent of the incoming flux, an approximation that is valid in the low-photons-count regime of the \highz \Lya forest, see][]{Eilers+2018}, convolving the spectrum with a Gaussian kernel, and finally re-binning the spectrum to the appropriate spectral resolution. The parameters used in the procedure vary as we match them to those of the dataset employed in the comparison.

In order to characterise the \highz \Lya transmission, we follow the same procedure used in \citet[][itself adapted from \citenp{Gnedin+2017}]{Garaldi+2019croc}. In short, we identify regions of \Lya transmission (spikes) and characterise them by a height ($h_\mathrm{s}$, corresponding to the maximum normalized transmitted flux), and a width ($w_\mathrm{s}$, defined as the simply-connected set of pixels that have $f \geq \alpha h_\mathrm{s}$,  where $\alpha$ is a tuneable parameter that we set to $\alpha = 0.5$ in this work). To complement this analysis we quantify the distribution of low-flux regions using the dark gaps (dg) statistics \citep[\eg][]{Paschos&Norman2005,Fan+2006,Gallerani+2006,Gallerani+2008}, defined as contiguous regions where the (normalised) transmitted flux is below some threshold. Once the latter is fixed, dark gaps are characterised uniquely by their length ($L_\mathrm{dg}$).

We stress here that, unlike many other numerical studies of the EoR (\eg \citenp{Becker+2015}, \citenp{Bosman+2018}, and \citenp{Eilers+2019} using a uniform UVBG, \citenp{Davies+2016}, \citenp{Kulkarni+2017} and \citenp{DAloisio+2018} using semi-analytical frameworks with multiple tunable parameters, and \citenp{Kulkarni+2018}, \citenp{Keating+18, Keating+2020} using post-processing RT simulations) we do not tune our simulations to explicitly match \Lya or post-reionization IGM properties. 
In fact, the only tunable parameter of the \thesan runs, \ie the stellar escape fraction, is set by requiring a late reionization history (see \paperI). 
Therefore, our predictions are directly and unequivocally linked to the physical modelling (including -- as it is virtually always the case -- free parameters tuned against some observations) employed in the simulations, that can consequently be tested in a more stringent way. 
We note here that, while tuning the stellar escape fraction to obtain a `late' reionization history surely affects the post-reionization IGM, such a connection is not trivial, in particular because \thesan employs a much more sophisticated star formation and feedback model with respect to the original works suggesting that cosmic reionization is completed at $z \lesssim 5.5$ \citep[\ie][]{Kulkarni+2018, Keating+18, Keating+2020}. 

With the exception of hydrogen and helium, whose ionization states are self-consistently tracked in the simulations, we compute the ionization state of other elements using \cloudy version 17.02 \citep{cloudy2017}.
Following previous works \citep[\eg][]{Bird+2015, Nelson+18} we employ the code in the single-zone mode (\ie assuming that the gas density and temperature are constant across a single resolution element of the simulation) and iterate to equilibrium, including both photo- and collisional ionization. 
We include the radiation background of \citet[][including its 2011 update]{Faucher-Giguere+09} and account for gas self-shielding using the prescription from \citet{Rahmati+2013}. We decided to proceed in this way, despite having self-consistent RT in the simulation, because of the broad frequency bins tracked by the code and relatively-narrow frequency range covered in comparison with the one required by \cloudy. 
We do not include induced processes \citep{Wiersma+2009} and use solar abundances from \citet{Grevesse+2010}. We compute the abundances of metal ionization states as a function of hydrogen number density $n_\mathrm{H}$, gas temperature $T$, and  redshift. In particular, we sample the following ranges: $-7.0 < \log( n_\mathrm{H} ) < 4.0$, $0 < \log(T) < 8.5$ and $5.5 < z < 10$ on a Cartesian grid. We note that the metallicity dependence is minimal since we are just computing the \textit{relative} abundance of each ionization state compared to the the neutral one for a given species. 
This procedure provides us the abundance $x_{i,j}$ of the $j$-th ion of the species $i$ on a 3-dimensional Cartesian grid, which we then interpolate (via spline interpolation) to each combination of $(n_\mathrm{H}, T, z)$ extracted from the simulation. Therefore, the number density of a given ion species can be obtained simply as $n_{i,j} = n_{i} \, x_{i,j}$, where $n_{i}$ is the number density of the metal species $i$, which is self-consistently tracked in \thesan for $9$ metal species (see Sec. \ref{sec:simulations}). 
This accurate forward modelling of the simulations allows us to faithfully study the properties of the \highz IGM, as presented in the following sections.

\subsection{Comparison with other approaches}
\label{sec:comparison}
Before presenting the results concerning the IGM--galaxy connection, we take a moment to compare our methodology to a number of different works focusing on the EoR. We refer the reader to \paperI for a more thorough discussion and to \citet{galform_review} for an overview of numerical simulations of galaxy formation. Over the years, a multitude of approaches have been developed. Many of them \citep[\eg][]{Keating+18, Keating+2020, Kulkarni+2017, Hassan+2021} rely on completely de-coupling the radiation transport from the hydrodynamics and galaxy formation, treating the former in post-processing. While this significantly reduces their computational cost, and hence enables the exploration of larger volumes of the Universe, it is then not possible to capture the radiation feedback on galaxies and on the gas in the CGM and IGM. The latter is crucial to properly characterise the connection and interplay between galactic and cosmological scales, which is one of the primary goals of \thesan and the main focus of this paper.

In recent times, semi-analytical models like RSAGE \citep{SAGE, RSAGE, RSAGE2} and MERAXES \citep{MERAXES} have attempted to include radiation feedback without the computational burden of full radiation-hydrodynamical simulations. However, this comes at the price of an approximate calculation of both the galaxy properties and the radiation field. 

Finally, fully-coupled radiation-hydrodynamical simulations of cosmic reionization have become available in recent times, often focusing on different selected aspects of the \highz Universe, a choice forced by the extreme computational cost of these calculations. For instance, SPHINX \citep{SPHINX} targets small scales, striving to resolve all the relevant photon sinks (hence removing the need for a stellar escape fraction) at the cost of relatively small box sizes. This is ideal for studying the internal properties of reionization sources, but the missing large scale coverage compromises the connection to cosmic reionization in a representative manner. 
The opposite approach is taken by the different iterations of the Cosmic Dawn project \citep{CoDa,CoDaII}, which simulates a significant volume of the universe but with coarse resolution (mainly as a consequence of the Cartesian grid employed). These calculations are well-suited to investigate IGM properties on cosmological scales (as \eg the 21cm radiation topology), but lack the resolution needed to produce realistic galaxies and, hence, meaningfully study the IGM -- galaxy connection. Finally, the CROC \citep{CROC} simulation suite is the closest one in terms of box size and resolution to \thesan. As we show in \paperI, however, high-mass galaxies in CROC deviate from the stellar-to-halo-mass relation inferred by  observations and other cosmological simulations. As we will argue in Sec.~\ref{sec:igm-gal-connetion}, this may be the culprit behind the inability of these simulations to reproduce the observed modulation of \Lya flux with galaxy distance \citep[][]{Garaldi+2019croc}, hence casting doubt on whether the IGM -- galaxy correlation can be predicted faithfully. A further difference is that CROC is  calibrated at high redshift, while \thesan is mostly calibrated at low redshift.

\section{Properties of the IGM}
\label{sec:IGM}

\begin{figure*}
\includegraphics[width=\textwidth]{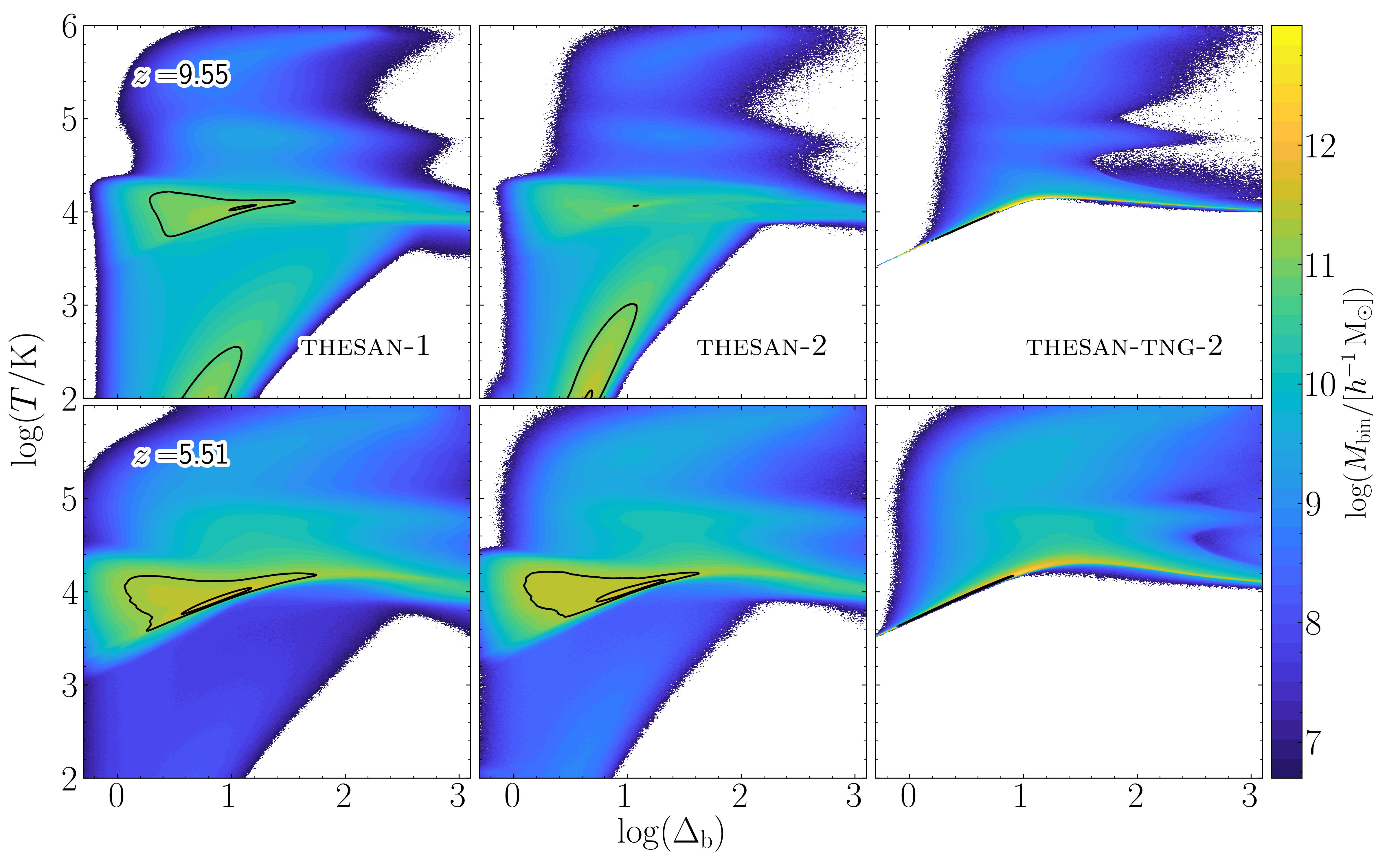}
\caption{Temperature-baryon overdensity phase space of the gas in the \thesanone (left), \thesantwo (center) and \thesantng (right) runs during the initial ($x_\ion{H}{I} \gtrsim 0.9$, $z\sim 10$, top) and final ($x_\ion{H}{I} \lesssim 0.01$, $z\sim5.5$, bottom) phases of reionization. The color-coding scales logarithmically with the mass in each pixel. Note that, in order to highlight the differences in the intergalactic medium, we plot only the most relevant part of the phase space. It can clearly be seen that the inclusion of self-consistent radiation transport and non-equilibrium thermo-chemistry greatly alter the properties of the low-density warm IGM, providing a significantly more realistic description than through the use of an uniform UVBG.}
\label{fig:phase_space}
\end{figure*}

We begin our analysis of the high-redshift IGM in the \thesan simulations focusing on its global properties. 
In order to provide an overview of the \thesan runs employed, we show in Fig.~\ref{fig:z_reion} the reionization redshift ($z_\mathrm{reion}$) in a slice through the different simulation boxes (whose name is reported in the top left corner of each panel). The largest panel shows our main fiducial run (\thesanone). 
In order to compute $z_\mathrm{reion}$ we employ the so-called Cartesian outputs, which save at high time cadence a subset of gas properties on a Cartesian grid (see \paperI for details). 
For each pixel, we determine the local reionization history, which typically shows multiple periods of full reionization ($x_\mathrm{HII} \geq 0.99$) interspaced by periods of complete neutrality ($x_\mathrm{HII} \leq 0.01$). We therefore define $z_\mathrm{reion}$ as the minimum redshift such that $x_\mathrm{HII}(\Tilde{z}) \geq 0.99$ for each $\Tilde{z} < z_\mathrm{reion}$ (\ie the last time the threshold value $x_\mathrm{HI} = 0.99$ is crossed from below). 
From a simple visual inspection, it can be appreciated that the reionization in \thesan proceeds inside-out, starting from the largest structures, which reionize their surroundings at $z \gtrsim 10$ (green to blue colors). Only much later do ionizing photons reach the lower-density IGM (yellow to red colors). The patchy nature of reionization is also evident from the Figure.
Finally, the densest structures (\eg individual galaxies and filaments) remain neutral until the final redshift of the simulations because of their high density and --~consequently~-- large recombination rates. In the Figure, they stand out as dark red structures embedded in blue patches. 
While counter-intuitive, there is broad consensus that individual galaxies are simultaneously the sources of photons and the only structures in the Universe to remain neutral well beyond the end of the EoR. This is made possible by the multi-phase structure of their ISM and CGM, which combines high densities that boost the (hydrogen) recombination rate, keeping the overall \ion{H}{II} fraction low, and highly-ionised channels that allow the escape of ionizing radiation in their surroundings \citep{Wise&Cen2009}. As such, we caution the reader that the details of the ionization state of these structures may depend on the specific ISM model employed in the simulations. 

Some features of the different physical models can be readily appreciated by comparing the panels in Fig.~\ref{fig:z_reion}. For instance, the bottom left panel displays how the IGM in \thesanlow is reionized significantly earlier than in the other runs, as highlighted by the overall more yellow colors. 
Moreover, comparing it with the \thesantwo and \thesanhigh simulations, differences in the sources of radiation become apparent. These runs exhibit a smoother $z_\mathrm{reion}$ in the IGM, as a consequence of the fewer, larger ionization sources driving large coherent ionization fronts. On the contrary, \thesanlow has the most inhomogeneous IGM. Nevertheless, the location of the earliest reionized patches is the same in all runs, showing that even when photons are forced to escape only from small galaxies, their clustering in overdense regions can produce an effect roughly comparable to our fiducial model.

By construction, $z_\mathrm{reion}$ condenses the information on the evolution of gas into a single number at each location. As such, it does not provide a straightforward insight on the IGM condition at any given redshift. 
Therefore, we complement it with Fig.~\ref{fig:lic}, where we show the evolution in \thesanone of the ionising photons density and flux along a slice (of width equal to $20$\% of the box size). 
In particular, each panel shows (at a different redshift, reported in the top right corner) the linear integral convolution \citep[LIC][]{cabral1993} of the photon density (acting as a background texture) and the (average) photon flux along the slice plane (providing the vector field). In order to better visualise the photon flux, we have added to the texture a small Gaussian noise (with zero mean). 
The color scale is the same in all panels. Hence, the varying hue of the picture is simply a reflection of the rapid evolution of the ionising background towards the end of the EoR. 
The Figure shows how galaxies and galaxy groups act as powerful sources of ionising photons, that travel approximately linearly until they are absorbed in a radiation sink.

We provide a more quantitative view of the global properties of the IGM in the top panels of Fig.~\ref{fig:igm_prop}, where we show as a function of redshift the volume-averaged hydrogen neutral fraction (left panel) and temperature at mean density ($T_0$, central panel). 
These panels reproduce and expands on figures 4, 5 and 6 of \paperI, and are included in order to provide a self-contained discussion. 
In the right panel, we show the CMB optical depth ($\tau_\mathrm{CMB}$) due to the integrated Thomson scattering of its photons off of free electrons --~mainly produced by reionization~-- as a function of the uppermost redshift considered in its calculation. 
Below the final redshift  $z_\mathrm{fin} = 5.5$ of the simulation we assume for this calculation that (i) hydrogen is fully ionized, (ii) helium is singly ionized  at $3 < z \leq 5.5$ and doubly ionized at $z\leq3$.  
In all cases, we report a collection of constraints from the literature. 
The reionization history of the \thesan runs is consistent with the bulk of observations, derived from the evolution of \Lya emitters and the number density of Lyman break galaxies \citep{Ota+2008, Ono+2012, Pentericci+2014, Choudhury+2015, Tilvi+2014, Mesinger+2015}, from the dark pixel statistics \citep{McGreer+2011, McGreer+2015, Lu+2020}, from GRB afterglow \citep{Totani+2006, Chornock+2013}, from the Doppler widths of \Lya absorption lines in the quasar near zones \citep{Bolton+2012}, from the quasar damping wing \citep{Mortlock+2011, Schroeder+2013, Greig+2017, Greig+2019, Wang+2020}, from CMB modeling \citep{Robertson+2013}, from the Gunn-Peterson optical depth \citep{Fan+2006, Davies+2018, Yang+2020, XQR30}, from the angular correlation function of \Lya emitters \citep{Sobacchi+Mesinger2015}, from the rest-frame UV continuum of galaxies \citep{Schenker+2014}, from the detection of \Lya emission in Lyman break galaxies \citep{Mason+2018, Mason+2019, Hoag+2019, Jung+20}, from a combination of \Lya luminosity function, clustering and line profile \citep{Ouchi+2010} and from the \Lya visibility (\citenp{Dijkstra+2011}. This is by construction, as we explicitly tune the stellar (unresolved) escape fraction of the simulations to obtain a realistic reionization history. We note here that all runs in the \thesan suite except \thesanlow follow a late reionization model, necessary to account for the $z\sim5.5$ opacity of the \Lya forest \citep[see][and the discussion in Section~\ref{sec:Lya}]{Kulkarni+2018, Keating+2020}. We refer the reader to \paperI for a more in-depth discussion.
The (very scarce at $z\gtrsim5.5$) constraints on $T_0$ are derived from the Doppler widths of \Lya absorption lines in the quasar near zones \citep{Bolton+2010, Bolton+2012}, from the \Lya forest power spectrum \citep{Walther+2019}, and from the Doppler widths of \Lya transmission regions \citep{Gaikwad+2020}. These constraints are mutually inconsistent (as they would require an unphysically rapid cooling of the IGM). The \thesan runs all lie between the observed values, with a mild dependence of the maximum $T_0$ on the duration of the reionization process \citep[see][and \paperI for a more detailed discussion]{DAloisio+2019}. Because of its earlier reionization, the IGM temperature peaks earlier in \thesanlow compared to the other runs.
Finally, the CMB optical depth predicted by the \thesan runs agrees very well with the estimation from \citet{Planck2018cosmo}. Only \thesanlow\ features too large of a $\tau_\mathrm{CMB}$, again as a consequence of its earlier reionization. However, recent re-analysis of the \citet{Planck2018cosmo} data by \citet{Pagano+2020} and \citet{deBelsunce+2021} found larger values of $\tau_\mathrm{CMB}$. While \thesanone is consistent with (although on the low end of) these two estimates, many of the other \thesan runs are not.

In the bottom panels of Fig.~\ref{fig:igm_prop} we investigate the effect of the box size on the global IGM properties, providing indications on the relevance of sampling variance in this study. 
For the sake of clarity, we only investigate this in the \thesanone run. In particular, we have divided the simulation box in two sets of $N_\mathrm{sub}=8$ and $N_\mathrm{sub}=64$ sub-boxes. As references, an individual sub-box in the latter set has a linear size twice as large as the fiducial SPHINX simulation \citep{SPHINX}. Within each set, the union of the sub-boxes covers the entire volume. 
We compute the redshift-dependent values of $x_\ion{H}{I}$ and $T$ within each sub-box, and plot their scatter around the full-box average as a shaded region in the bottom panels of Fig.~\ref{fig:igm_prop}. It can be seen that, while the scatter obtained using $64$ sub-boxes is significant, it is almost-entirely suppressed when $N_\mathrm{sub}=8$. 
While conclusive statement about the impact of sample variance cannot be made without running a larger equal-resolution box (which is currently unfeasible because of numerical limitations), these results are encouraging. In fact, each sub-box of the $N_\mathrm{sub}=8$ set already shows a global evolution hardly distinguishable from the full-box one, indicating that the role of the box volume is marginal for volumes  $V \gtrsim (L_\mathrm{box}/2)^3$. Therefore, we expect sample variance to play an equally-marginal role in the result presented in this work. 

Reassured by this finding, we continue our investigation of the IGM properties by showing in Figure~\ref{fig:phase_space} the joint distribution (with colors reflecting the mass in each pixel) of gas temperature and baryon overdensity. 
The latter is defined $\Delta_\mathrm{b} \equiv \rho_\mathrm{b} / \bar{\rho}_\mathrm{b}$, where $\rho_\mathrm{b}$ is the baryon density (that in the IGM coincides with the gas density) and $\bar{\rho}_\mathrm{b}$ is its average value. 
The top row of the Figure displays the gas state at $z = 9.5$, when the reionization process is still in its early stages, while the bottom row shows the gas state reached at the tail-end of the EoR ($z = 5.5$), where most of the gas in the simulation has already been fully ionized. 
The Figure also presents a comparison between \thesanone (our flagship simulation, left column), \thesantwo (employing the same physical model at  $8\times$ lower mass resolution, central column), and \thesantng (employing the original IllustrisTNG model to evolve the same initial conditions as \thesantwo, right column). 
Finally, each panel shows black contours encompassing the central $68$\%, $95$\% and $99.5$\% of the data. Before presenting our interpretation of these results, we note here that the range of baryonic overdensities and temperatures reached in our simulations is much larger than shown in the panels. 
In particular, in the \thesan simulations temperatures down to $5$~K and up to $\sim 10^8$~K are found, as well as baryonic overdensities up to $\Delta_\mathrm{b} \sim 10^8$. 
However, we choose to focus here on the most relevant region of phase space for the IGM --~the main topic of this paper~-- during and after reionization. 
In addition, at large densities, the gas follows an effective equation of state \citep{Springel&Hernquist03}, which suppresses most of the excursion in temperature.

A first striking difference can be appreciated comparing the right column with the others. 
The original IllustrisTNG model does not self-consistently track radiation, as \thesan does, but rather employs a spatially-uniform UV background \citep{Faucher-Giguere+09}. 
This approach results in a very tight temperature-density relation with little redshift evolution. 
Conversely, a more accurate treatment of radiation (including photo-heating and non-equilibrium thermo-chemistry) allows us to fully capture the inhomogeneous process of reionization, which can be seen in the left and central columns as a large distribution of temperatures associated with a single density value. 
This is the result of the overlap of individual regions ionized at different times. 
Each of them establishes its own temperature-density relation as a consequence of photo-heating, which then evolves because of gas cooling \citep[via adiabatic expansion and Compton cooling; \eg][]{Trac+08,Furlanetto&Oh09}. 
We note here that, while at $\log \Delta \lesssim 1$ the RT is responsible for the differences, the right-hand part of the diagram is ruled by hydrodynamical and galactic effects. 
Therefore, the junction of these two regimes can be faithfully captured only by combining self-consistent RT with an accurate galaxy-formation model, as in \thesan. 
For instance, employing post-processing RT, even on finely-spaced outputs from hydrodynamical simulations, results in a mismatch in the $\log \Delta \gtrsim 1$ region of the phase space \citep{Keating+18}. 
Employing a spatially-uniform UV background (as routinely done in large hydrodynamical simulations and as done in \thesantng)  entirely suppresses the low-temperature gas found during reionization when RT is self-consistently implemented. 
(We note here that, additionally, the \thesan simulations employ a non-equilibrium solver for the thermal state of the gas, unlike the IllustrisTNG simulations.) 
The high-temperature portion of this diagram (corresponding to $T \gtrsim 10^{4.5}$~K) is remarkably similar in all the runs presented in the Figure, since gas is brought to such a thermal state by energetic feedback processes originating from galactic physics, which is unchanged in the three runs.
Finally, we note that there are small differences in the IGM structure at $z=9.55$ between \thesanone and \thesantwo. 
These are a consequence of the slightly delayed reionization history in the latter (see Fig.~\ref{fig:igm_prop}), which results in a difference in the phase space when compared at the same redshift.

\begin{figure}
\includegraphics[width=\columnwidth]{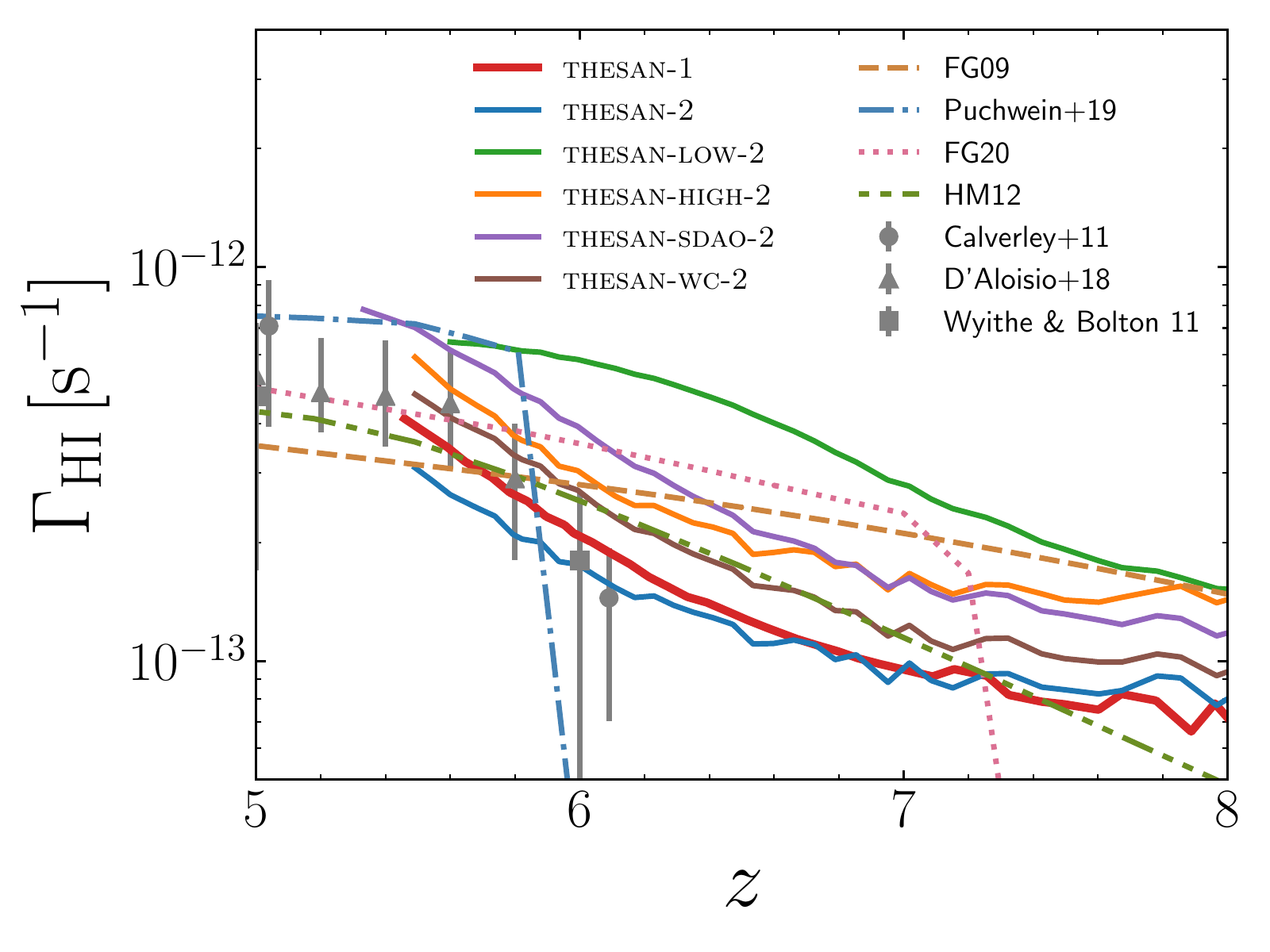}
\caption{Photo-ionization rate for \textsc{HI} computed in ionized regions (\ie where $x_\mathrm{HII} \geq 0.5$). Solid lines show the results extracted from different \thesan runs, the dashed, dot-dashed, dotted and double-dashed lines report the predictions from the UVBG models of \citet[][including its update in December 2011]{Faucher-Giguere+09}, \citet{Puchwein+19}, \citet{Faucher-Giguere20}, and \citet{Haardt&Madau12}, respectively. Symbols show measurements from \citet{Calverley+11, DAloisio+2018, Wyithe&Bolton11}.}
\label{fig:gamma}
\end{figure}

\begin{figure}
\includegraphics[width=\columnwidth]{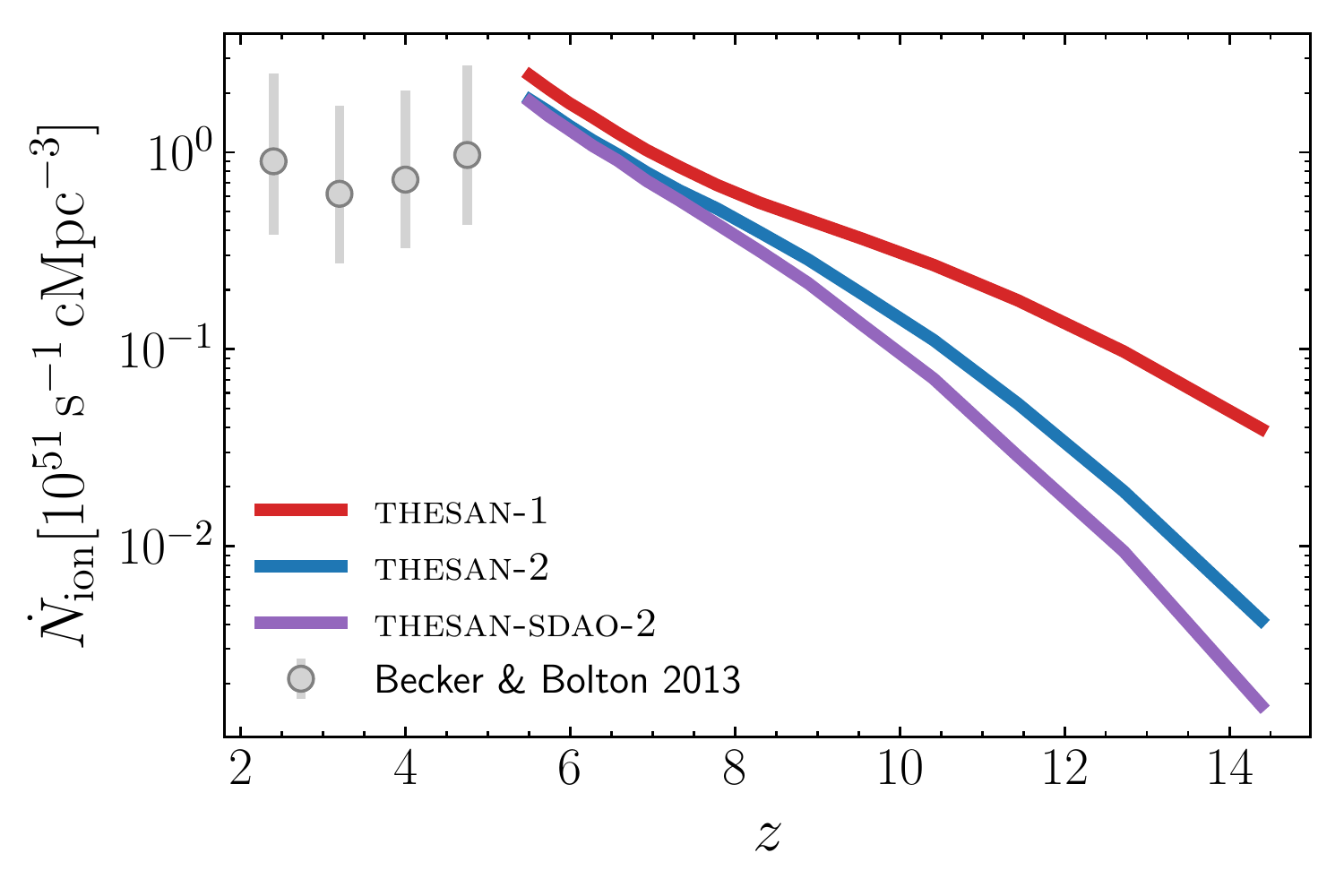}
\caption{Evolution of the photon emission rate from stars in the \thesanone, \thesantwo and \thesansdao runs, compared to observations of \citet{Becker&Bolton2013}. Since this quantity depends only on the stellar population in the simulations, the \thesan runs not shown here overlap almost perfectly with \thesantwo.}
\label{fig:dotN_vs_z}
\end{figure}

\begin{figure}
\includegraphics[width=\columnwidth]{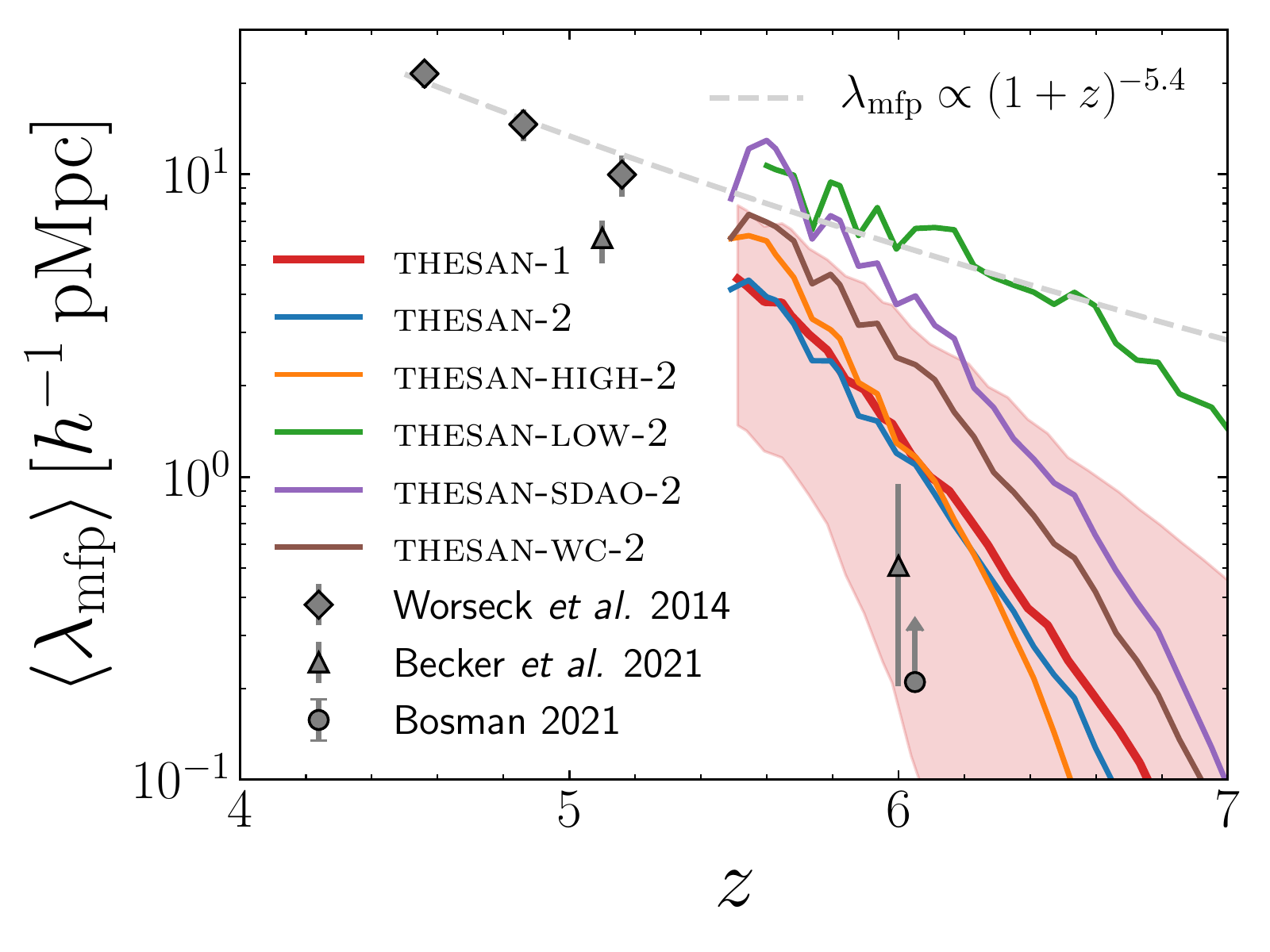}
\caption{Mean free path of ionizing photons as a function of redshift, compared to observations from \citet{Worseck+2014}, \citet{Becker+2021} and \citet[][displaced by $\Delta z = 0.05$ for visual clarity]{Bosman2021}. The dashed line shows the fit to $z < 5.5$ data reported in \citet{Worseck+2014}. \thesanone reproduces the recently observed rapid growth of the mean free path, as a consequence of its late reionization history. Medium resolution simulations show a similar evolution, except for \thesanlow, which features a much earlier reionization history, close to the classical picture.}
\label{fig:mfp}
\end{figure}

The density and temperature of the IGM crucially depends on the position- and frequency-dependent radiation field. 
We exploit here the self-consistent RT to investigate the accuracy of available UVBG models. 
We begin by showing in Figure~\ref{fig:gamma} the \HI photo-ionization rate $\Gamma_\HI$, computed in ionized regions (defined as gas cells with neutral hydrogen fraction $x_\HI < 0.5$) as a function of redshift. 
The solid lines in the Figure show the outcome from a selection of the \thesan simulations, while symbols report values observed by \citet{Calverley+11}, \citet{DAloisio+2018}, and \citet{Wyithe&Bolton11}.
Almost all \thesan runs are compatible with available data. However, it appears clear that the simulated curves have a shallower evolution than observed. In fact, while data seem to rise by a factor $\sim10$ between $5.5 \leq z \leq 6$ and then flatten, in simulations $\Gamma_\ion{H}{I}$ steadily grows since $z\sim8$ and shows no sign of stagnated growth by $z=5.5$. 

Interestingly, the only \thesan run largely not matching the available data is \thesanlow. 
This is a consequence of the earlier completion of the reionization process in this case. 
We can try to approximately account for the different reionization history of \thesanlow by replacing the original curve $\Gamma(z)$ with $\Gamma(\Tilde{z})$, where $\Tilde{z}$ is the redshift that matches the reionization history of \thesanlow with the one of \thesanone, \ie such that $x^\thesanlow_\ion{H}{I}(\Tilde{z}) = x^\thesanone_\ion{H}{I}(z)$. This corresponds approximately to a delay of $\Delta z \sim 0.8$ and brings \thesanlow in better agreement with (although still partially higher than) \thesanhigh and \thesansdao. 
These three curves lie consistently above the ones from \thesanone and \thesantwo, and partially the one corresponding to \thesanwc, and the reason can be found in the different number of stellar sources effectively contributing to the EoR. 
In fact, while in the fiducial physical model every star particle injects photons in the local environment, we suppress the photon escape from low-mass galaxies in \thesanlow and from high-mass ones in \thesanhigh. 
Similarly, the \thesansdao run has modified initial conditions that suppress small scale power, reducing the number of small galaxies. 
Hence, for their reionization history to be consistent with available data, the stellar escape fraction $f_\mathrm{esc}$ needs to be higher in these models, in order to counter-balance the reduction of ionizing sources, effectively producing a larger photon density (and hence $\Gamma_\mathrm{HI}$) in the ionized regions. 

Another significant conclusion that can be extracted from the analysis of Figure~\ref{fig:gamma} is that \thesanone and \thesantwo show very similar values of $\Gamma_\mathrm{HI}$ throughout most of the simulation, indicating good convergence of the two models. 
Only below $z\lesssim6$ does \thesantwo start to lag behind \thesanone, and this is reflected in its slightly-delayed reionization history. 
Additionally, when compared to values inferred from observations, the fiducial physical model fits the data best among the \thesan runs, as the other are generally higher than the data at $z\gtrsim5.8$. 
Nevertheless, the observed $\Gamma_\HI$ shows a steep increase (by almost an order of magnitude) between $z=6.1$ and $z=5.5$, and a subsequent flattening at lower redshift. Conversely, the evolution in all the \thesan runs is significantly slower and its slope does not become shallower down to $z\sim5.5$. This may be indicative of a reionization history that is too extended in our simulation. 
There are, however, other explanations. For instance, the number of haloes above $M_\mathrm{thr} = 10^{10} \, \Msun$ increases with time. Therefore, the number of haloes contributing to reionization in \thesanlow decreases at lower redshift, explaining the flattening in this model. 
Finally, we compare the simulated evolution of $\Gamma_\HI$ with a selection of UVBG models available in the literature. 
The dashed, dot-dashed, dotted and double-dashed lines in the Figure report the predictions from \citet[][FG09, including its update in December 2011]{Faucher-Giguere+09}, \citet{Puchwein+19}, \citet[][FG20]{Faucher-Giguere20}, and \citet[][HM12]{Haardt&Madau12}, respectively. Surprisingly, the model that most closely follows the results from the \thesanone and \thesantwo runs is the one from HM12, while FG09 more closely resembles the results from \thesanhigh and \thesansdao. 

Before moving on to the characterisation of the mean free path of ionizing photons, we show in Fig.~\ref{fig:dotN_vs_z} the evolution of the volume-averaged ionizing photon emission rate in \thesanone, \thesantwo and \thesansdao, compared to available measurements from \citet{Becker&Bolton2013}. Note that other medium resolution runs are virtually identical to \thesantwo in this regard, since their star-formation history is only very mildly affected by the different radiation fields. 
At very high redshift, \thesanone produces a significantly larger number of ionizing photons than \thesantwo, thanks to its higher resolution able to resolve smaller objects (see also figure 12 in \paperI). At $z\lesssim8$, these two runs show a similar rate of injection of ionizing photons. 
Similarly, the suppression of some small-scale modes in the \thesansdao initial conditions reduces the number of ionizing photons emitted at $z\gtrsim8$ with respect to \thesantwo. 
By redshift $z\sim8$, however, $\dot{N}_\mathrm{ion}$ is similar in all runs. 
If extrapolated to $z \leq 5$, the value would be slightly larger than observed by \citet{Becker&Bolton2013}, showing that an inversion in this trend is needed to match the observed constraints. 
In \citet{Kulkarni2019} and \citet{Keating+2020} such a sharp decrease in $\dot{N}_\mathrm{ion}$ had to be assumed in order to match the mean \Lya flux evolution, although these same authors claim it is very challenging to find a plausible physical interpretation. 
\citet{Ocvirk+2021} found that a strict upper limit on the temperature of the gas allowed to form stars can reproduce such a drop.\footnote{We note here that the authors conclude that the drop in $\dot{N}$ is a consequence of the photo-ionization feedback from percolating reionized bubbles. However, the different distribution of stars produced by the two star-formation prescriptions likely creates significant differences in the stellar and supernova feedback on the gas, which may produce effects on star formation at least as strong as photo-ionization feedback. }
Our simulations adopt a similar approach, preventing gas with temperatures above the one predicted by its effective equation of state (\eg as a consequence of heat injection from feedback processes) from forming stars. Despite this, we do not see any sign of decrease in $\dot{N}_\mathrm{ion}$ at $z\lesssim6$. 
The reason for this difference may lie in the fact that feedback processes typically heat the gas to temperatures that are orders of magnitude higher than the one employed by \citet{Ocvirk+2021} as an upper temperature threshold.

\subsection{Mean free path evolution}
\label{sec:mfp}

\begin{figure*}
\includegraphics[width=\textwidth]{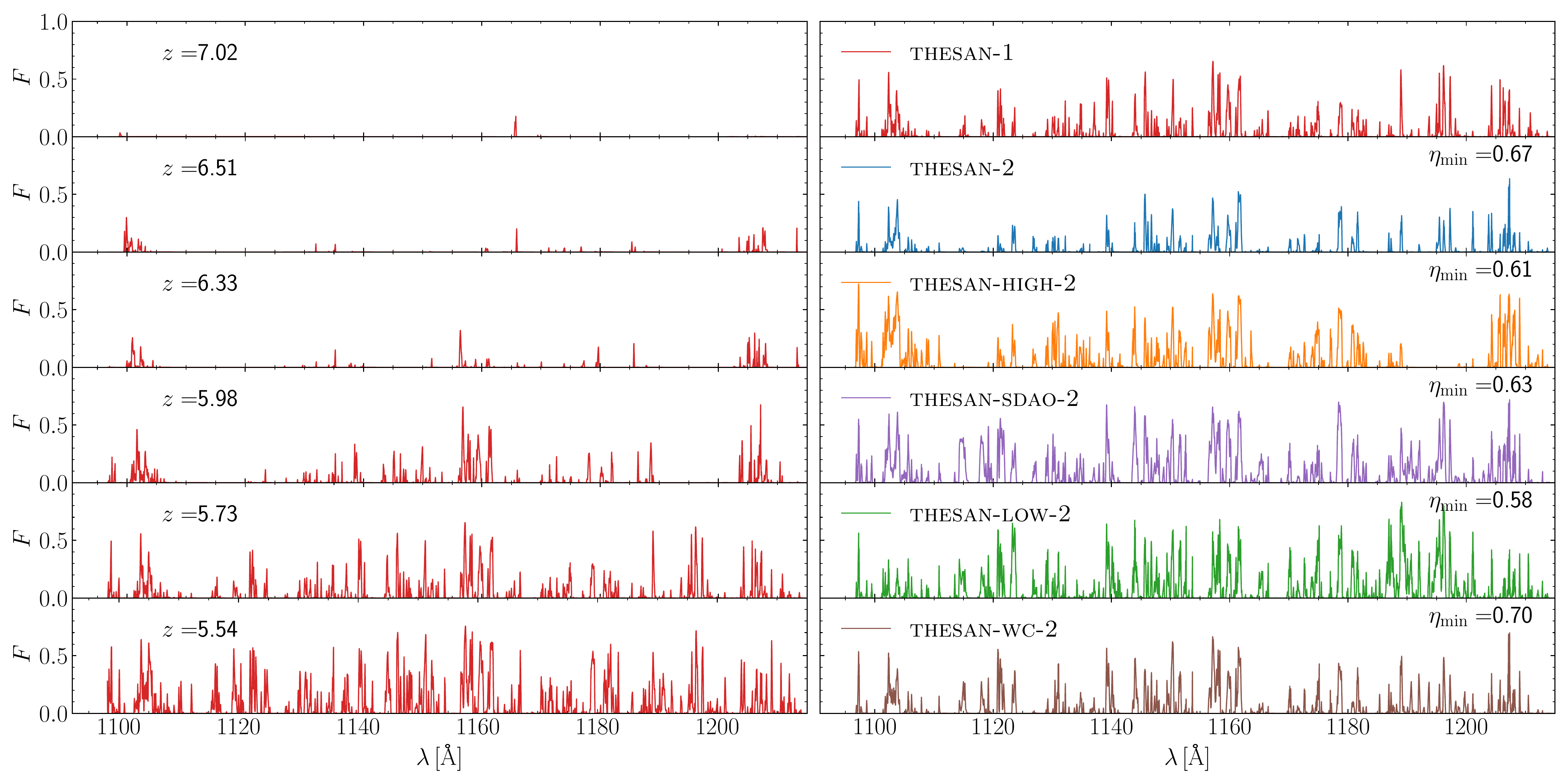}
\caption{Example of forward modeling synthetic spectra in the \thesan suite. The left column shows the evolution with redshift of the same line of sight in the \thesanone flagship run. The right column displays the different \Lya spectra obtained at the same redshift ($z=5.73$) in a selection of \thesan runs. The vertical axis shows the normalized transmitted flux while the horizontal one reports the rest-frame wavelength. In the top right corner of the right-hand-side panels, we report the overlapping index $\eta_\mathrm{min}$ of the sightlines in the run shown and in \thesanone.}
\label{fig:samelos}
\end{figure*}

Very recent measurements of the mean free path of ionizing photons ($\mfp$) from the attenuation of \highz quasar spectra point to a sharp drop in this quantity between $z\sim5$ and $z\sim6$ \citep{Becker+2021, Bosman2021}. 
This newly-observed $\mfp$ lies significantly below the value at the same redshift predicted by extrapolating measurements at $4 \lesssim z \lesssim 5$ \citep{Worseck+2014}. 
We compare these new measurements with the predictions from the \thesan suite in Figure~\ref{fig:mfp}, where we also show the distribution of the central 68\% of ionizing photons free paths computed from the simulations in \thesanone (other \thesan runs have similar scatter). 
We stress that the simulations were \textit{completed} before such measurements were made public. 
In order to compute $\mfp$ from the simulated IGM, we follow \citet{Rahmati+2018} and extract random (both in origin and direction) skewers through the simulated volume and determine the mean free path, $\mfp$, from the following definition:
\begin{equation}
    \int_0^{\mfp} \dd s \, \sigma_{912} \, n_\mathrm{HI} = 1 \, ,
\end{equation}
where $\sigma_{912}$ is the cross-section for photons with wavelength $\lambda = 912 \, \text{\AA}$ and $\dd s$ is the line-of-sight distance. 
The predicted evolution of $\mfp$ in our fiducial model is in excellent agreement with the measurements from \citet{Becker+2021} and the lower limit from \citet{Bosman2021} at $z\sim6$ and appear to approach the available data at $z\sim5$. 
The same holds true for the \thesantwo and \thesanhigh runs, which are in very good agreement with the results of our flagship run. The agreement between \thesanone and \thesantwo appears here somewhat fortuitous, since these two runs have both a different reionization history and different resolution (\ie resolve small-scale structures to a different degree). 
This can be seen comparing them with \thesanwc, which has a reionization history very close to the one of \thesanone and shows --~on average~-- a larger mean free path. 
Additional evidence in support of this interpretation comes from the inspection of the evolution of $\mfp$ with the ionized fraction (instead of redshift, removing the dependence on the reionization history, not shown for the sake of brevity), which appears indistinguishable in \thesantwo and \thesanwc, but offset toward lower values in \thesanone.
This also demonstrates that the \thesan simulations are not completely converged in their predictions of $\mfp$. 
It appears likely that, in order to match the median $\mfp$ in \thesanone with the central value obtained by \citet{Becker+2021}, a slightly delayed reionization history is required.

Differently from the other runs, \thesanlow shows a much larger mean free path at all redshifts, as a consequence of its significantly earlier reionization history. Note that once reionization is completed in this model, its $\mfp$ settles on the extrapolation of the \citet{Worseck+2014} observations, suggesting that in \thesanlow the post-reionization mean free path is compatible with the observed one. 
Most of the difference can be ascribed to the early reionization history of this run. However, even after accounting for such effects (see above), there are residual differences among \thesantwo, \thesanlow and \thesanhigh. In particular, the latter has a larger mean free path than our fiducial physical model, while the photon free path in \thesanlow is systematically lower than in \thesantwo \citep[similarly to the recent results of][]{Cain+2021}. 
Finally, \thesansdao exhibits values of the mean free path that are a few times larger than the observed ones at $z\sim6$, and slightly larger than \thesanwc, again as a consequence of the slightly earlier completion of reionization (we have checked this is the correct interpretation by inspecting $\mfp (x_\mathrm{HI})$). It is interesting to note that, even with the strong suppression of certain (low) mass haloes in the sDAO initial conditions compared to the CDM ones, the mean free path is not significantly affected (once the slightly different reionization histories are accounted for).

Recently, \citet{Cain+2021} employed a coarse-grained RT simulation coupled with higher-resolution small-volume runs to approximately resolve small-scale structures in large volumes, and concluded that: (i) the rapid evolution in the mean free path observed by \citet{Becker+2021} favours a reionization driven by small, high-escape-fraction galaxies, and (ii) additional sinks are needed beyond those predicted by high-resolution numerical simulations to match the central value obtained by \citet{Becker+2021}. 
The results presented here partially support the first conclusion, showing that a reionization epoch driven by the smallest (largest) galaxies produces a shorter (larger) mean free path (at fixed reionization history). However, the ionizing photon budget in \thesanone is dominated by galaxies residing in haloes with mass $M_\mathrm{halo} \geq 10^{10} \, \Msun$ at $z \leq 8$ (see Figure~12 in \paperI). 
Additionally, our fiducial model is capable of recovering the $\mfp$ evolution in a way similar to \citet{Cain+2021} 
in a self-consistent RHD simulation framework. 
We note here that \thesanone resolves haloes with mass $M_\mathrm{halo} \gtrsim 10^8 \, \Msun$ (this value is $8$ times larger for the lower-resolution runs). These are smaller haloes than the smaller sub-resolution sinks in the fiducial configuration of \citet{Cain+2021}.  Additionally, our simulations can resolve sinks of significantly smaller mass if they are gas dominated, thanks to the better gas mass resolution with respect to that of the DM. 

In \citet{Davies+2021}, the recent measurements of the mean free path at $z\sim6$ were combined with the coeval dark pixel fraction to infer a cumulative output of ionizing photons per baryon $n_{\gamma/\mathrm{b}} = 6.1^{\tiny +11}_{\tiny -2.4}$, a number significantly larger than typically assumed. Moreover, these high values require an escape fraction $\gtrsim 20$\% across the entire galaxy population. In \thesanone, we find that $n_{\gamma/\mathrm{b}} = 6.31$ photons per baryon are needed to reionize the IGM, in agreement with the value quoted above. In order to compute this number, we consider the total ionizing output from stars and include an effective escape fraction of photons into the IGM, which includes both the unresolved escape from the birth cloud, modelled through the $\fesc$ parameter, and the resolved absorption from the ISM and CGM of simulated galaxies, resulting in an effective escape fraction of order $\sim 10$\% for most of the simulation (see the right panel of Figure~17 in \paperI and the relative discussion therein).

Studying the IGM is very challenging, especially at \highz, because its low density renders its emission very dim, leaving absorption studies as the only available approach. The main absorption line studied in this context is \Lya, which we investigate next.

\section{The high-redshift Ly\texorpdfstring{$\balpha$}{α} forest}
\label{sec:Lya}

We now move to an in-depth analysis of one of the main observables of the EoR, the \Lya transition of neutral hydrogen. We have described in Section~\ref{sec:spectra} the process for the production of synthetic absorption spectra that are employed here, and we show in Fig.~\ref{fig:samelos} an example of the resulting spectra. 
In particular, the left column shows the evolution of the transmitted flux (as a function of rest-frame wavelength) along the same line of sight extracted from the \thesanone run. As expected, the number of transmission regions grows with redshift. 
In the right column we show the same LOS at $z=5.73$ in the different runs employed in this Paper (and summarised in Table~\ref{table:simulations}). While the position and characteristics of the transmission features are overall similar, small differences among the different runs exist, which warrant the further in-depth analysis that we carry out in this Section. 
Before proceeding further, we quantify these differences in the right column of Fig.~\ref{fig:samelos} using the overlapping coefficient
\begin{equation}
    \eta_\mathrm{min} (f,g) \equiv \int \min[f(x), g(x)] \, \mathrm{d}x\ \Big/ \int \mathrm{d}x \, ,
\end{equation}
where $f$ and $g$ are two arbitrary functions with unit integral and the integration runs over their entire (common) support. $\eta_\mathrm{min}$ is unity in case of perfect overlap and vanishes in case at least one function is vanishing at each common support point. In our case, we fix $f$ to be the (appropriately normalised) flux in the \thesanone run (top right panel) and $g$ to be the (appropriately normalised) flux along the same sightlines in the different run shown. We report the value of this overlap coefficient in the top right corner of the right-column panels. 

\begin{figure}
\includegraphics[width=\columnwidth]{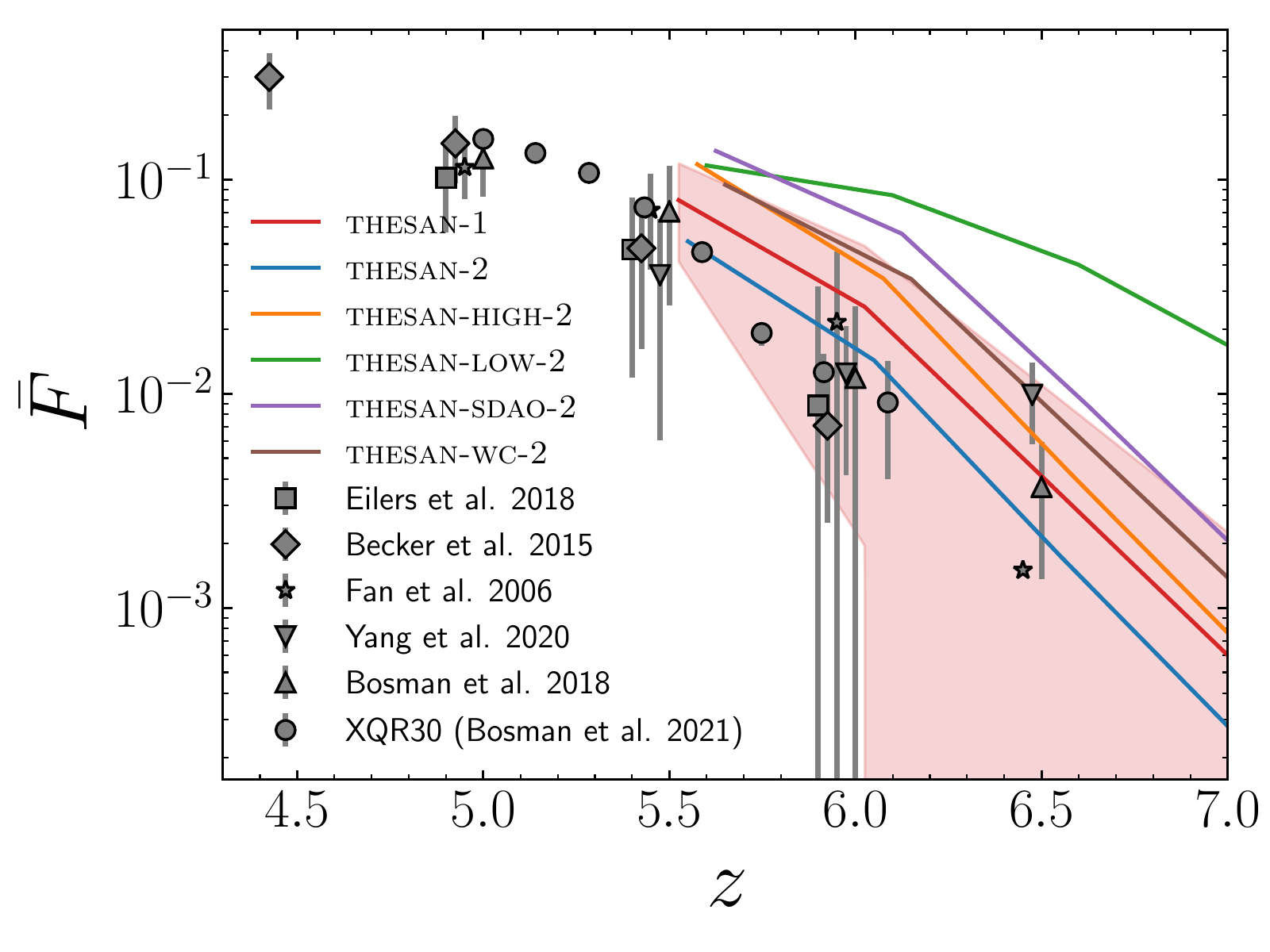}
\caption{Mean transmitted flux in the \Lya forest as a function of redshift for a selection of \thesan runs, compared to the observations of \citet{Fan+2006}, \citet{Becker+2015}, \citet{Bosman+2018,XQR30},  \citet{Eilers+2018}, and \citet{Yang+2020}. For clarity, we show the central $68$\% of the data for \thesanone only as a shaded region. The other simulations exhibit a similar scatter.}
\label{fig:mean_flux}
\end{figure}

The simplest measure of the high-redshift \Lya transmission is the average transmitted (normalised) flux $\meanf$, shown in Figure~\ref{fig:mean_flux}. 
Symbols report values computed from observed \highz quasar spectra \citep{Fan+2006,Becker+2015,Bosman+2018,Eilers+2018,Yang+2020,XQR30}, while solid lines show the values computed from the \thesan suite. For the sake of clarity, we show the central $68$\% of the data with a shaded region only for \thesanone, as the other runs show a similar scatter. 
This simple plot already reveals interesting patterns. First, similarly to the cases discussed in Section~\ref{sec:IGM}, the runs employing our fiducial physical model (\ie \thesanone and \thesantwo) are most consistent with data. 
Despite the extreme sightline-to-sightline variation in $F$ and the consequent scatter, the various simulations show consistent differences that remain across all investigated redshifts. 
In particular, all but \thesanlow show a virtually identical redshift evolution with different amplitudes. Taking \thesanone as a reference, we now explore such trends. 
Sightlines in \thesantwo typically show a slightly lower transmitted flux, consistent with its somewhat delayed reionization history, which is a consequence of the smaller number of low-mass galaxies resolved in this simulation (see \eg Fig.~\ref{fig:dotN_vs_z}). 
Conversely, \thesanhigh exhibits an enhancement in $\meanf$ with respect to \thesanone. 
This can be readily explained by the fact that in this run ionizing photons are only emitted by large galaxies, while the stellar escape fraction is increased to complete reionization at approximately the same time as \thesanone. 
Therefore, in this run ionized regions are typically larger and have higher ionized fractions, overall producing an enhancement of the transmitted flux. 
In other words, since we explicitly adjusted $f_\mathrm{esc}$ to maintain the volume-averaged ionized fraction approximately the same, the concentration of photons around a small number of bright sources produces a larger transmission when compared to the same IGM location in the \thesanone run. 
This can be seen in Figure~\ref{fig:samelos}, comparing the first and third panels from the top in the right column. 
The former shows slightly more transmission regions than the latter (notice \eg the second group of transmission peaks from the right, which is present in \thesanone but absent in \thesanhigh) because of the larger number of ionized regions, but where both spectra have non-zero transmission, \thesanhigh shows typically a larger flux (appreciable \eg in the leftmost group of high-transmission regions). 
We will explore the feasibility of using the statistics of \Lya transmission regions to constrain the sources of reionization in a forthcoming work.

Figure~\ref{fig:mean_flux} additionally shows that there exists an offset between \thesansdao and \thesanone, similar to the one described above but larger in size. In the former run, in fact, the values of $\meanf$ are significantly larger than in the latter at all redshifts. 
This is also reflected in its earlier completion of reionization. 
An extreme case is represented by \thesanlow, which shows a much larger value of $\meanf$ that additionally evolves more slowly with redshift and remains almost constant below $z < 6$, in stark contrast with available observations. 
It is important to note here that, unlike in previous discussions, a na\"ive translation of the curve to lower redshifts to account for the earlier reionization in \thesanlow does not bring this run into agreement with observations. 
In fact, in the latter, the value of $\meanf$ increases more than linearly with redshift, while \thesanlow exhibits the opposite behaviour. 
The reason for this lies in the fact that the contribution of small galaxies (\ie with DM-halo mass $M_\mathrm{h} \leq 10^{10} \, \Msun$), the only ones injecting ionizing photons in the IGM in this model, decreases significantly towards the end of the EoR, ultimately stagnating reionization. This can quantitatively be seen by looking at the contribution to the global star formation-rate density from galaxies residing in haloes of different mass. This quantity is shown in Figure 12 of \paperI for \eg the \thesantwo run, whose galaxy and stellar populations are virtually identical to those in \thesanlow.

Finally, our flagship simulation \thesanone, while consistent with available measurements of $\meanf$, appears to lie slightly higher than the latter at $z \lesssim 6$. 
However, the extreme difficulties linked to measuring the \Lya transition at such high redshifts, the patchy nature of reionization, as well as the limited availability of sightlines, render these measurements subject to large uncertainties. 
Hence, we believe that our fiducial model provides a realistic description of \Lya transmission during the reionization epoch.

\subsection{Effective optical depth distribution}
\label{sec:tau_eff}

\begin{figure*}
\includegraphics[width=\textwidth]{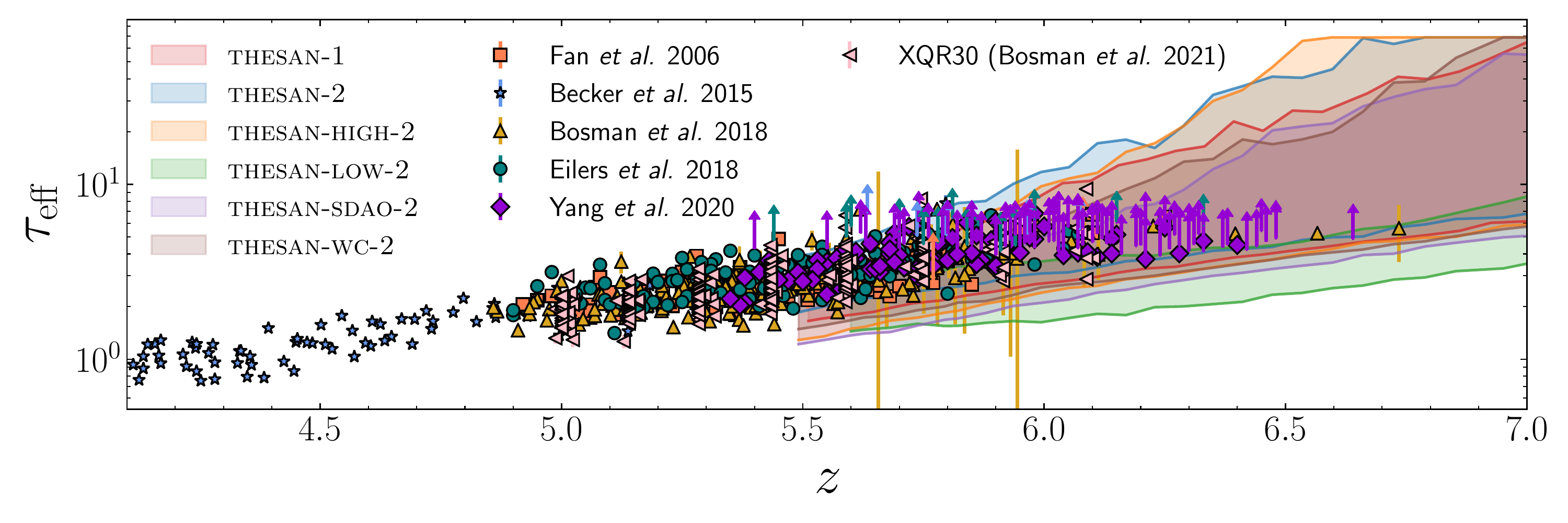}
\caption{Evolution of the effective \Lya optical depth ($\taueff$) in different redshift bins, averaged over spectral chunks of length $50\,\hMpc$. The shaded regions show the central 95\% of the data computed from synthetic spectra in a selection of \thesan runs, while symbols report individual observations of \citet{Fan+2006}, \citet{Becker+2015}, \citet{Bosman+2018, XQR30}, \citet{Eilers+2018}, and \citet{Yang+2020}.}
\label{fig:tau_evol_individual}
\end{figure*}

The trends described above are also reflected in the evolution of the effective optical depth $\taueff \equiv - \ln \avgf$ (where $\avgf$ is the mean flux in spectral segments of length $50 \, \hMpc$), as shown in Figure~\ref{fig:tau_evol_individual}. Shaded regions display the central 95\% of the distributions extracted from a selection of the \thesan simulations, while symbols and arrows report measurements and lower limits --~respectively~-- from \citet{Fan+2006}, \citet{Becker+2015}, \citet{Bosman+2018, XQR30},  \citet{Eilers+2018}, and \citet{Yang+2020}. 
With the exception of \thesanlow, which exhibits significantly lower optical depth with respect to the other runs at all redshift, the predictions for all other \thesan runs are very consistent with each other. 
Additionally, they match the observed $\taueff$ remarkably well at all redshifts. 
At the highest redshifts, most observations provide only lower limits, since the flux in the \Lya forest is below the detection threshold. 
This results in an apparently stalled evolution of $\taueff$, which our simulations predict to simply be an artefact of the increasing difficulty of detecting the diminishing transmitted flux in observations.

\begin{figure*}
\includegraphics[width=\textwidth]{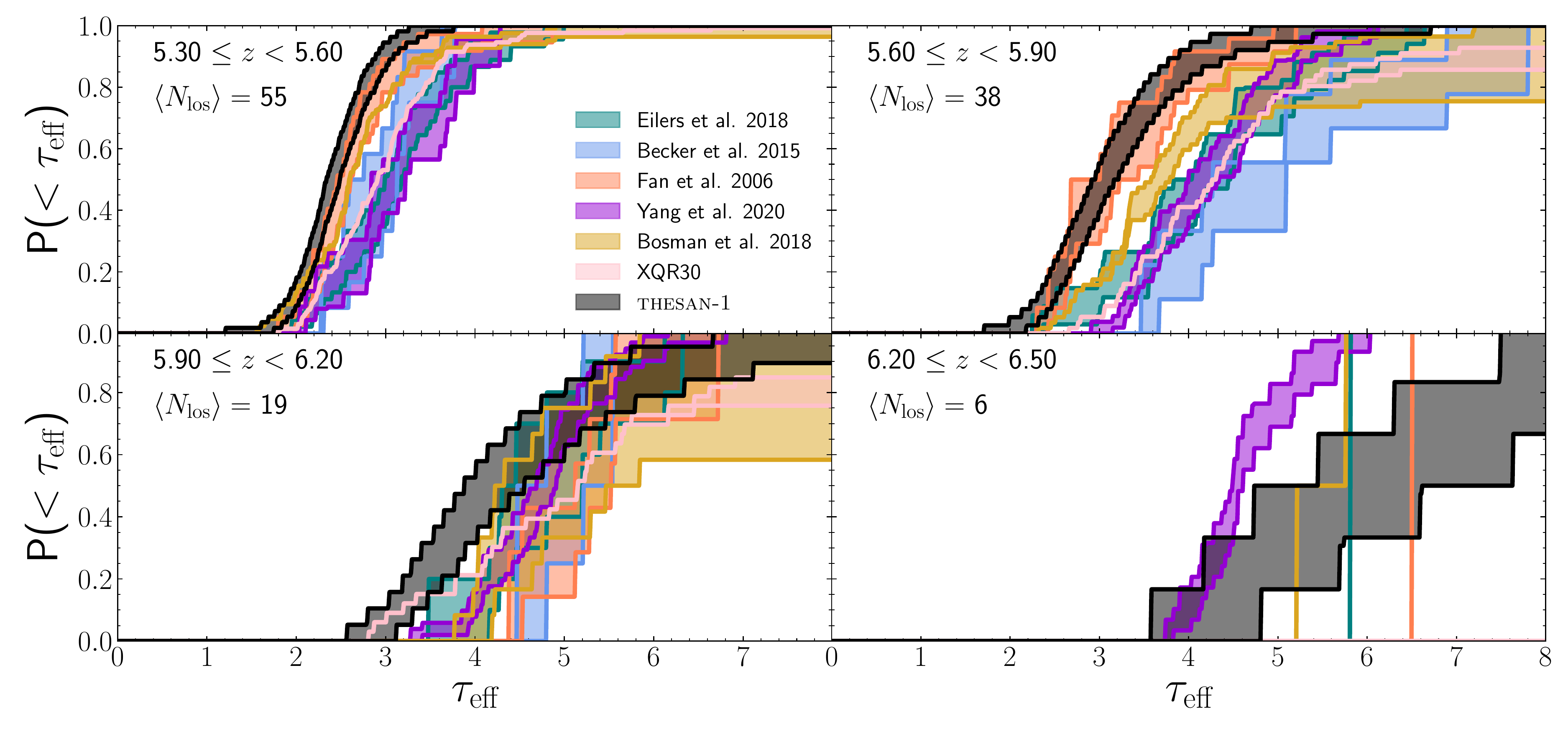}
\caption{Cumulative distribution function of effective \Lya optical depths in different redshift bins, averaged over spectral chunks of length $50\,\hMpc$, and compared to the observations of \citet{Fan+2006}, \citet{Becker+2015}, \citet{Bosman+2018, XQR30}, \citet{Eilers+2018}, and \citet{Yang+2020}. The ranges are computed by bootstrapping (with $500$ samples) the original distribution and showing the $15$th and $85$th percentile of the samples, except for the \citet{Bosman+2018} sample, where we follow their optimistic-pessimistic approach (see the text for an explanation).}
\label{fig:tau_cdf}
\end{figure*}

With this caveat in mind, we now move to a more quantitative analysis of the cumulative distribution function (CDF) of effective optical depths at a fixed redshift in Figure~\ref{fig:tau_cdf}. It shows the cumulative distribution of $\taueff$ in four different redshift bins, centred on $z = 5.45, 5.75, 6.05, 6.45$ with a width of $\Delta z = 0.3$. 
The choice of these redshift bins is motivated by the requirement of having a sufficient number of observed sightlines in each bin to meaningfully compare with the simulated ones. 
We report their average number across the different datasets  
in the top left corner of each panel, together with the redshift covered by the bin. 
For each dataset, we estimate the distribution of CDF via bootstrapping, creating $500$ realizations of the sample and showing the central $68$\% of them.
In addition, only in the case of \thesanone, we force each bootstrapped sample to contain a number of sightlines equal to the average number in the observed dataset, in order to account for sample variance.
We account for the finite sensitivity of the observations by assuming that the intrinsic flux equals $\max(\avgf, 2\sigma_{\avgf})$, where $\sigma_{\avgf}$ is the error on $\avgf$.
A slightly different strategy is taken for the \citet{Bosman+2018} and XQR30 \citep{XQR30} datasets, in order to follow the approach of the original authors. In fact, they provide an `optimistic' and a `pessimistic' method for the handling of low-flux pixels. The former is identical to the procedure just described. The latter, instead, assumes the intrinsic flux is vanishing (\ie $\taueff \rightarrow \infty$). 
Finally, we do not attempt to model observational errors on the transmitted flux in this context, since they are different for each (and within each) dataset, preventing a fair comparison. 
However, \citet{Eilers+2018} showed that --~for \Lya~-- their effect is typically to slightly steepen the high-$\taueff$ end of the CDF at redshift $z\gtrsim6$. 
This is a consequence of the fact that whenever the flux is below twice the associated error, the latter is assumed while computing the optical depth. 
Hence, including error degrades the sensitivity towards high optical depth, effectively placing an upper limit on the values of $\taueff$ that can be obtained. 
We have checked that in our case this effect is negligible. 
With the exception of the highest-redshift bin, the results from \thesan agree with the observed distribution of optical depth, although they lie on the lower $\taueff$ side of the observed values. 
This may indicate that the Universe is reionized even later than what is predicted by \thesanone, or that our resolution is preventing us from resolving some of the densest structures, acting as sinks of photons. 

We can test these effects by comparing the predicted CDF in \thesanone, \thesantwo, and \thesanwc, in order to separately vary the resolution and the reionization history. 
We show such comparison in Figure~\ref{fig:tau_cdf_comparison}, where we display the CDF in a selection of \thesan runs at $z=5.5$ (top panel) and at the time when the volume-averaged hydrogen neutral fraction is $x_\mathrm{HI} = 0.02$ (bottom panel). In both cases we use redshift bins of width $\Delta z = 0.2$ centered on the redshift reported. For the sake of visual clarity, we do not report the range computed via bootstrap, but only the original CDFs. 
The top panel clearly illustrates that the effective optical depth distribution is primarily controlled by the reionization history. In fact, \thesanwc appears much more similar to \thesanone than \thesantwo by virtue of having a very similar reionization history to the former and despite having the same mass resolution and force softening as the latter. 
The small incongruity between the weak convergence run and our flagship simulation can be entirely explained by the small residual difference in their reionization histories (see Fig.~\ref{fig:igm_prop}), as shown in the bottom panel, where all curves except \thesanhigh and \thesanlow perfectly overlap. It is interesting to note that --~despite synchronizing the reionization histories~-- there are large differences in the distribution of $\taueff$ when galaxies of different masses are forced to single-handedly power the reionization process. (We have checked that this feature does not change when choosing different hydrogen neutral fractions.)
Therefore, it appears possible --~at least theoretically~-- to combine observations of $\taueff$ and $x_\mathrm{HI}$ to infer precious information on the sources primarily responsible for the reionization of the Universe. Additionally, the bottom panel of Figure~\ref{fig:tau_cdf_comparison} highlights again the excellent numerical convergence of the \thesan simulations with respect to the properties of the \highz IGM. Once the different reionization histories in \thesanone, \thesantwo, and \thesanwc (due to the slightly different star-formation histories and stellar escape fractions, see \paperI) are accounted for, the properties of the IGM appear indistinguishable. 

In the highest-redshift bin (\ie $6.2 \leq z \leq 6.5$) of Figure~\ref{fig:tau_cdf} the CDF of optical depth in \thesanone appears significantly different from the data from \citet{Yang+2020}, the only one with more than a single measurement at these redshifts. 
This seems to imply that \thesan predicts too much inhomogeneity in the IGM, therefore requiring an earlier reionization and/or \textit{fewer} resolved substructures. 
It appears very challenging to reconcile this requirement with what can be inferred at lower redshifts, unless the reionization history of the Universe is quite extended and shows little evolution between $z\sim6.5$ and $z\sim5.5$.
Since data in this redshift range originates from a single dataset we can meaningfully compare our simulations with, we have checked that forward modelling the errors in the synthetic spectra\footnote{
   We include errors on the mean transmitted flux in synthetic spectra by assigning to each spectral chunk of the simulated spectra an error $\sigma_{\avgf}$ randomly extracted from the errors in the \citet{Yang+2020} dataset in the same redshift bin. 
   $\taueff$ is then computed in the very same way as in observations. 
   We additionally check the effect of shifting the value of $\avgf$ by an amount sampled from a Gaussian distribution of zero mean and width equal to $\sigma_{\avgf}$, in order to mimic the effect of systematic changes in the mean flux due to such errors. We find that this additional step does not produce any appreciable difference in the optical depth distribution at this redshift.
}
 does not significantly alleviate the disagreement between \thesanone and the data from \citet{Yang+2020}.

\subsection{Characterisation of transmission regions}
\label{sec:dg_hp_wp}

\begin{figure}
\includegraphics[width=\columnwidth]{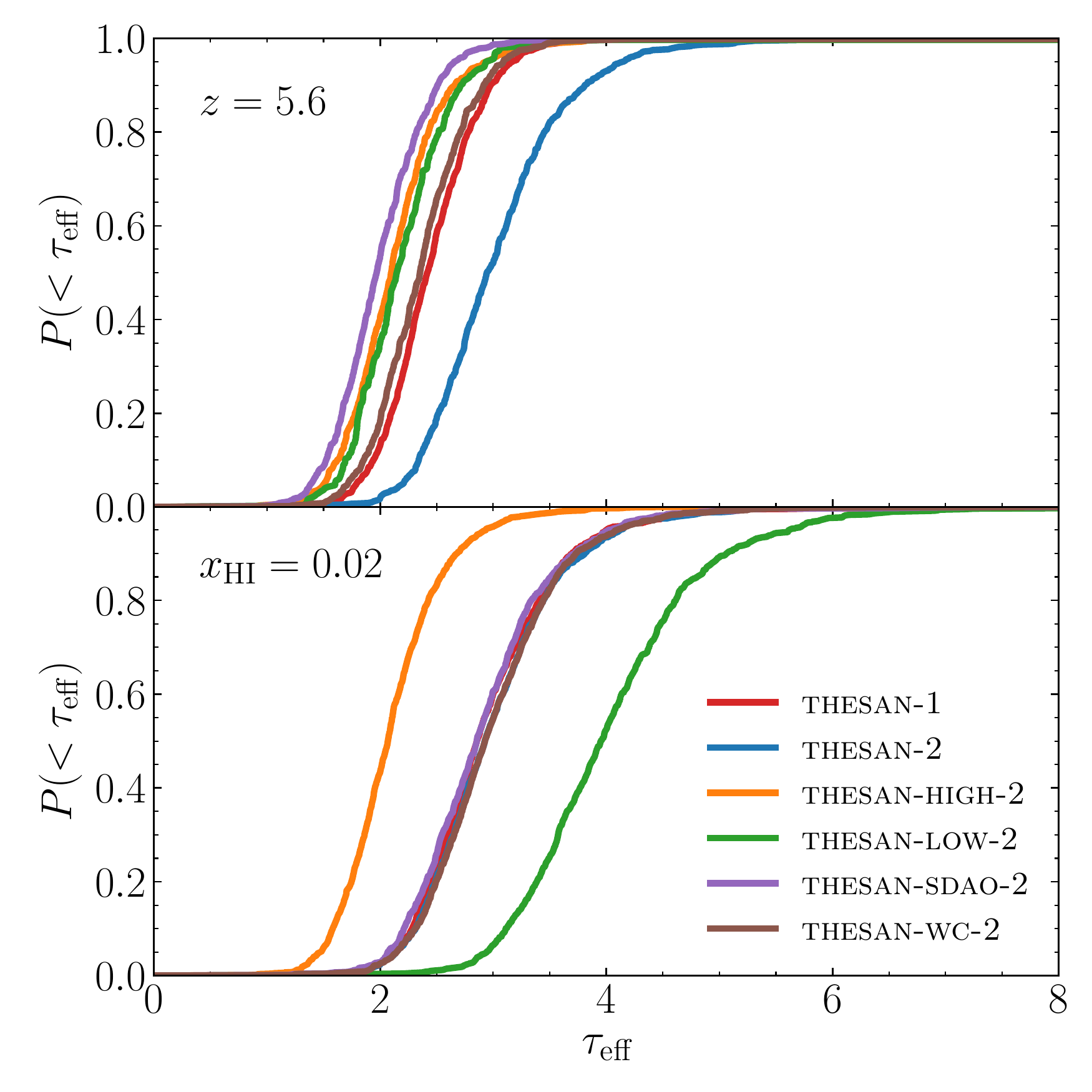}
\caption{Cumulative distribution function of effective \Lya optical depths at $z=5.6$ (top) and at the same volume-averaged hydrogen neutral fraction $x_\mathrm{HI} = 0.02$ (bottom) for a selection of \thesan runs.} 
\label{fig:tau_cdf_comparison}
\end{figure}

\begin{figure*}
\includegraphics[width=\textwidth]{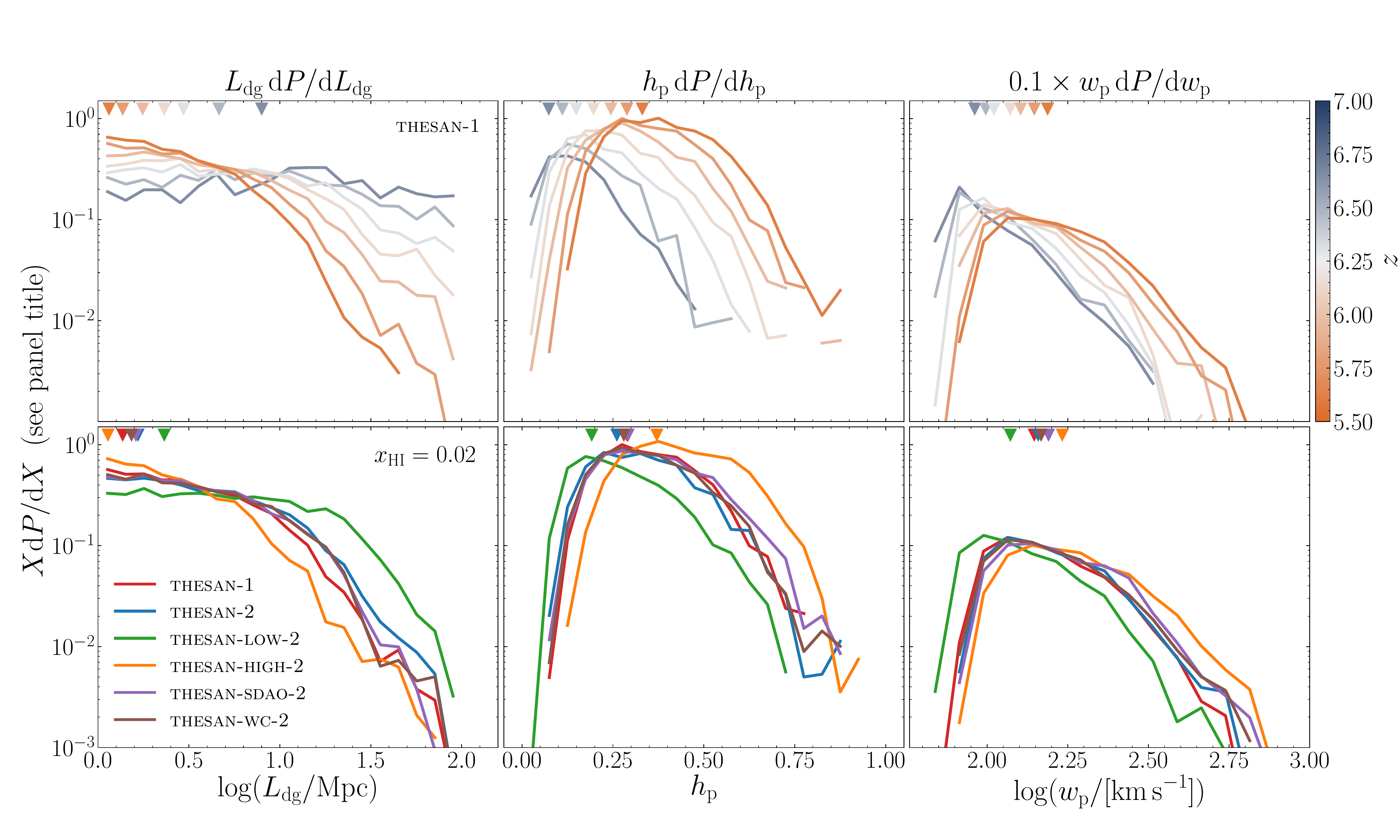}
\caption{\textit{Top}: Distribution of dark gaps lengths ($L_\mathrm{DG}$, left panel), peak heights ($h_\mathrm{p}$, central panel), and peak widths ($w_\mathrm{p}$, right panel) in the \Lya forest as a function of redshift in the \thesanone run. \textit{Bottom}: Same as above, but at fixed volume-averaged neutral fraction ($x_\mathrm{HI} = 0.02$) in a selection of \thesan runs. Triangles on the top part of the panels show the location of the mean of each distribution.}
\label{fig:DG_hp_wp}
\end{figure*}

The analysis performed in this Section focuses on the transmitted flux in the \Lya forest. Until now we have only considered its amplitude. In the following, we move to a deeper analysis, characterising the shape and distribution of the transmission regions. More specifically, we employ the dark gaps (dg) and peaks definition presented in Section~\ref{sec:spectra}. 
In Fig.~\ref{fig:DG_hp_wp} we show the distribution of dg lengths ($L_\mathrm{dg}$, left panel), peak heights ($h_\mathrm{p}$, central panel), and widths ($w_\mathrm{p}$, right panel) in the \thesanone simulation as a function of redshift (indicated by the line color, top row) and for the different \thesan models at the time when the volume-averaged neutral fraction is $x_\mathrm{HI} = 0.02$ (approximately $z\sim5.5$ for most models, except for \thesanlow, see Fig.~\ref{fig:igm_prop}). Specifically, we show the quantity $X \, \mathrm{d} P(X) / \mathrm{d}X$ for each feature $X = \{L_\mathrm{dg}, h_\mathrm{p}, w_\mathrm{p}\}$, where the probability density is computed as \citep{Gnedin+2017}:
\begin{equation}
    \frac{\mathrm{d} P(X)}{\mathrm{d}X} \equiv \frac{\Delta N (X, \Delta X)}{N_\mathrm{tot} \Delta X} \, ,
\end{equation}
with $\Delta N$ being the number of features in a bin centred on a given value $X$ and of width $\Delta X$, and $N_\mathrm{tot}$ is the total number of features detected. 
It should be noted that the length of dark gaps is limited to the length of synthetic spectra, which in turn is constrained by the box length. This only becomes relevant at $z\gtrsim6.5$, where the number of such long dark gaps is not negligible.
For this analysis, we convolve the spectra with a boxcar filter that effectively degrades their spectral resolution to $R=2000$, in order to mimic observations of these features \citep{Gnedin+2017}.

A first apparent feature visible from the top panels of Fig.~\ref{fig:DG_hp_wp} is the redshift dependence of the distributions. The length of dark gaps (left panel) shows a flat distribution at $z\sim7$, while at lower redshifts the number of long gaps is progressively suppressed as more highly-ionized regions emerge around the sources of radiation. By $z\sim6$, the probability of finding dark gaps larger than $10$ Mpc drops steeply. 
This highlights the possibility of using the tail of this distribution to constrain the timing of reionization, similarly to what has been done recently by \citet{Keating+2020} with an individual sightline containing  an exceptionally-long Gunn-Peterson trough. However, our results show that the occurrence of short gaps can also provide valuable information on the reionization history of the Universe, opening up the possibility of employing the full distribution to constrain the latter. 

Initial steps in this direction were taken in \citet{Gnedin+2017}, although in that work a single reionization history was investigated. 
Differences in the distribution of dark gap lengths can also arise also from the different source models for reionization. Therefore, in the bottom left panel of the figure we show the distribution in different \thesan runs at the same volume-averaged hydrogen neutral fraction $x_\mathrm{HI} = 0.02$. This removes differences arising from unequal reionization histories (although some residual differences may be found since we are now comparing distributions at slightly different redshifts, and therefore the density field in these runs is not exactly the same, despite originating from the same initial conditions). 
We have checked that the residual differences in the temperature evolution after the reionization history is controlled for do not play a major role. For instance, the evolution of $T(x_\mathrm{HI})$ in \thesanone and \thesanlow is practically indistinguishable, while they are well separated in the bottom panel of Fig.~\ref{fig:tau_cdf_comparison}. It is however desirable that these results are checked in greater detail, especially given their potential for constraining the sources of reionization.
All runs that share the same source model (\thesanone, \thesantwo, and \thesanwc) show indistinguishable distributions. Interestingly, the distribution computed from \thesansdao overlaps with the former ones, showing that this statistical measure is not able to distinguish \LCDM from the sDAO dark matter model. 
This is likely a consequence of the fact that the sDAO model differs from the \LCDM one only in the abundance of small dark matter haloes, therefore impacting mostly the initial phases of reionization, which are dominated by such objects (see \paperI), while becoming indistinguishable from the other runs once more massive objects start to dominate the ionizing photon budget. 
It should be noted, however, that the definition of dark gaps itself depends on a parameter (the flux threshold) that can change the sensitivity of $P(L_\mathrm{dg})$ to different physical quantities. 
Finally, \thesanlow and \thesanhigh show a somewhat counter-intuitive behaviour. The former shows more long gaps while the latter shows fewer, despite the former harbouring more ionizing sources than the latter. This behaviour stems from the fact that the small galaxies sourcing reionization in \thesanlow produce lower transmission with respect to \thesanhigh and, therefore, more of these peaks remain below the flux threshold used to define a dark gap in the former, but not in the latter, effectively joining together what in \thesanhigh are multiple adjacent gaps. This is also the reason why the number of short gaps is larger (smaller) in \thesanhigh (\thesanlow).

It is interesting to note that the shape of $P(L_\mathrm{dg})$ depends on redshift and the reionization model (\ie driven by low- or high-mass galaxies) in different ways. 
In particular, the main effect of the former is to split long dg into multiple, shorter ones, decreasing $P(L_\mathrm{dg} \gtrsim 10\,\mathrm{Mpc})$ while boosting $P(L_\mathrm{dg} \lesssim 5\,\mathrm{Mpc})$. Interestingly, the position of this break remains approximately fixed. 
Varying the source models, instead, moves the position of this break towards larger (smaller) lengths when only photons from low (high) mass galaxies reach the IGM. This is very important as it enables us to disentangle the two effects from observational data. 

The behaviour described is consistent with the results presented in the central and right panels of Fig.~\ref{fig:DG_hp_wp}. In particular, the distributions of both $h_\mathrm{p}$ and $w_\mathrm{p}$ move to larger values as reionization progresses, the IGM becomes more ionized and the ionized patches grow larger \citep[see][for a detailed discussion of how the peak shape depends on IGM properties]{Garaldi+2019croc}.
Once the reionization history is controlled for, \thesanlow exhibits peak heights that are, on average, smaller than in our fiducial physical model. Conversely, the transmission peaks in \thesanhigh are on average higher.
A similar trend can be seen in the peak widths, with the former (latter) model producing narrower (broader) peaks than in \thesanone.

\section{The IGM -- galaxy connection}
\label{sec:igm-gal-connetion}

\begin{figure*}
\includegraphics[width=\textwidth]{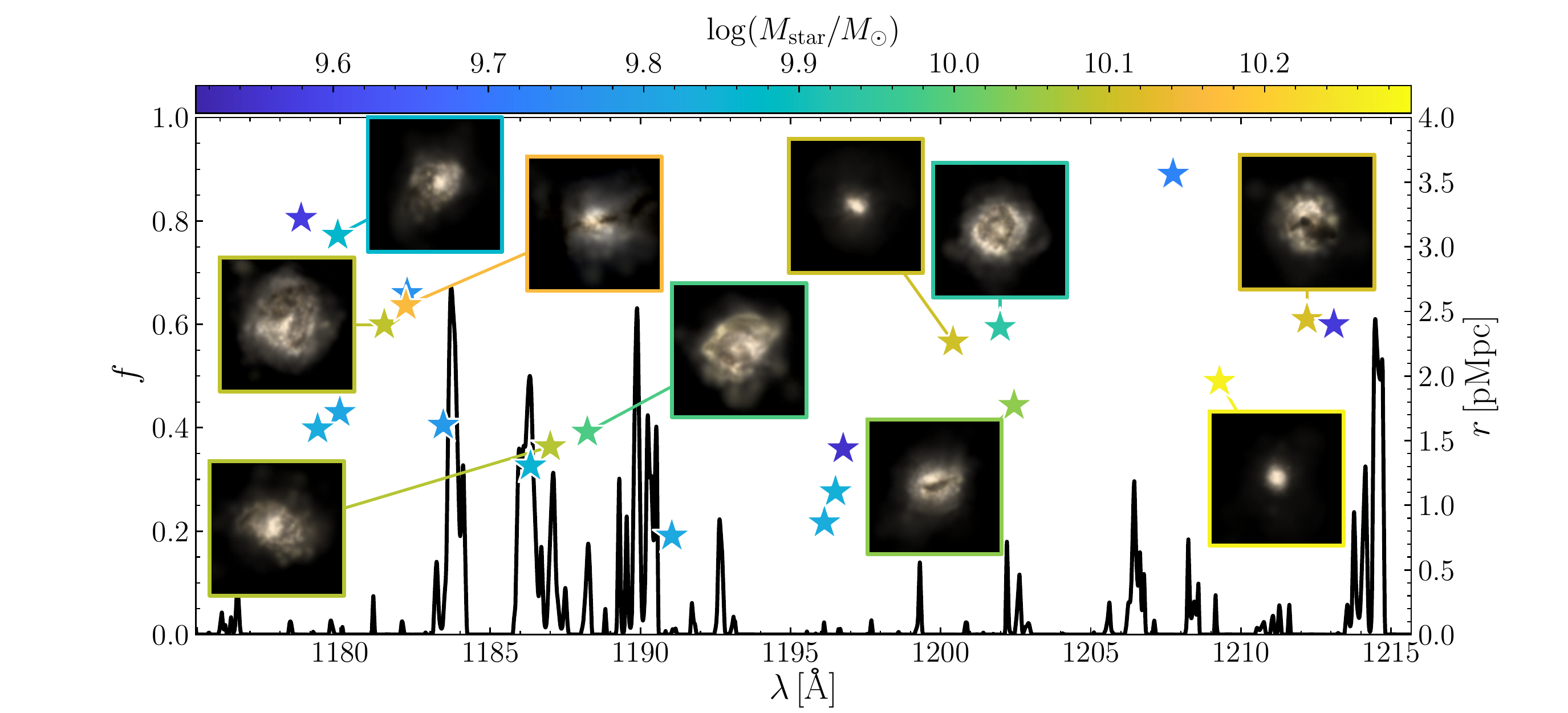}
\caption{Normalised \Lya transmitted flux along a sightline through the \thesanone box (left vertical scale). Star symbols show the
surrounding galaxies with stellar mass $M_\mathrm{star} \geq 10^{9.5} \, \Msun$, colour-coded according to the latter. Their vertical position reflects their distance from the sightline (right vertical scale). Insets show synthetic \jwst mock images of the ten galaxies in the panel with the largest $M_\mathrm{star}$, combining the F277W, F356W and F444W NIRCam wide filters and covering an area of $(10 \, \mathrm{pkpc})^2$.}
\label{fig:los_galaxies}
\end{figure*}

\begin{figure}
\includegraphics[width=\columnwidth]{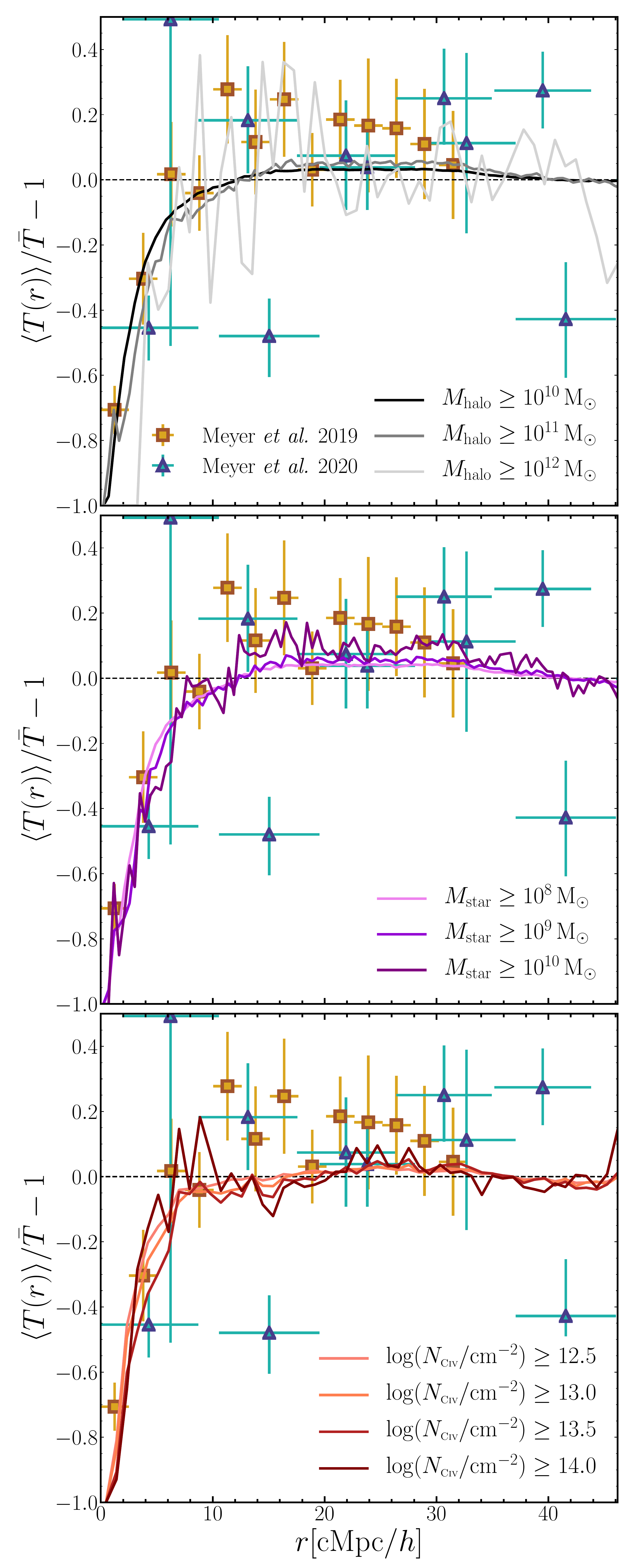}
\caption{Average transmission in the \Lya forest in the \thesanone run as a function of distance from nearby galaxies at $z=5.5$. Lines show the effect of considering only a sub-population of halos or galaxies while computing this quantity. Symbols show the measurements from \citet[][yellow squares]{Meyer+2019} and \citet[][cyan triangles]{Meyer+2020}.}
\label{fig:correlation-th1}
\end{figure}

\begin{figure}
\includegraphics[width=\columnwidth]{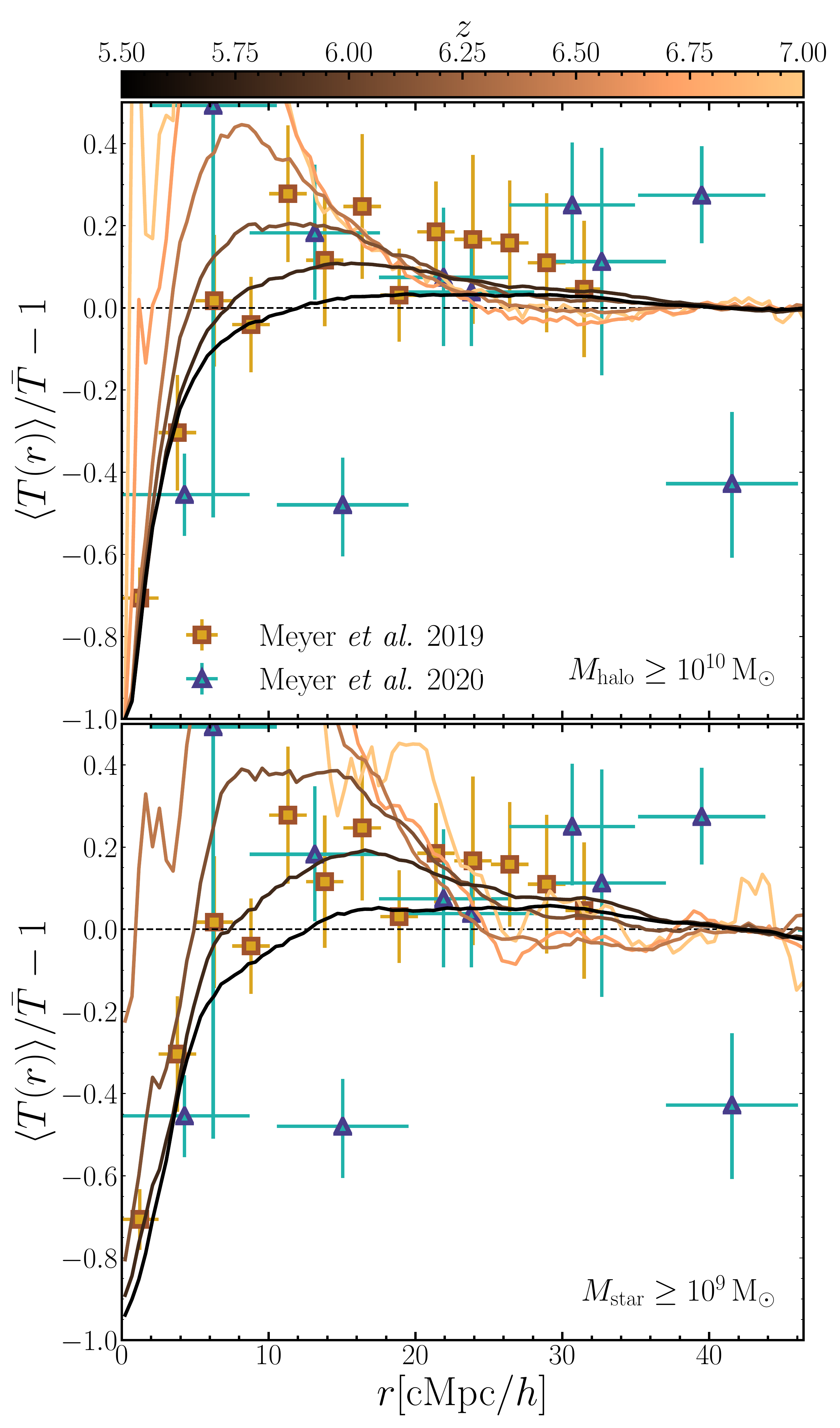}
\caption{As Fig.~\ref{fig:correlation-th1}, but now showing the redshift evolution for a single selection criterion of $M_\mathrm{halo} \geq 10^{10} \, M_\odot$ for the top panel and $M_\mathrm{star} \geq 10^9 \, M_\odot$ for the bottom one. }
\label{fig:correlation-evol}
\end{figure}

In the previous sections we have demonstrated that \thesan predicts very well the observed properties of the \highz IGM. In \paperI we have shown that our simulations produce a realistic galaxy population at $z \gtrsim 5.5$ (and we are confident the same is true at lower redshifts since we use the successful IllustrisTNG galaxy formation model). Hence, we investigate here the connection between the high-redshift galaxies and the IGM. 

We do so by using the dependence of the average transmissivity $T$ (\ie the ratio between the transmitted flux and the inferred continuum flux at the same location) in the \Lya forest at a distance $r$ from nearby galaxies as a probe of the connection between galaxies and the IGM at $z\geq5.5$. 
This test that was originally proposed and observationally performed by \citet{Kakiichi+2018}. 
Subsequently, additional observations, as well as a refinement of the approach itself, have been provided by \citet{Meyer+2019} and \citet{Meyer+2020}. 
In particular, \citet{Meyer+2019} have shown that at $4.5 \lesssim z \lesssim 6.3$ the transmitted flux displays a broad peak at intermediate distances to galaxies $ 10 \lesssim r / [\mathrm{cMpc}/h] \lesssim 30$, which has been interpreted as a proximity effect of the latter. In fact, under the assumption that galaxies provide the main contribution to the reionization process, their surroundings are expected to experience a larger-than-average density of ionizing photons, which entails a larger IGM transmission.\footnote{
  Notice that, even in the case QSOs were the dominant source of ionizing photons (which is not the case in \thesan, see \paperI), we expect the aforementioned flux excess to exist. In fact, QSOs are co-spatial with a subset of galaxies, therefore producing a flux enhancement around the latter. However, since QSOs are significantly more rare, we also expect the signal excess to be washed out when computing the correlation using all galaxies.}

On the theoretical side, however, this approach has been barely investigated, potentially because it simultaneously requires large simulated volumes and radiative transfer techniques to faithfully capture the IGM properties, as well as a realistic galaxy formation model, which in turn often requires high resolution. For these reasons, \thesan is ideally suited to study this quantity. 
\citet{Garaldi+2019croc} have investigated the ability of the state-of-the-art Cosmic Reionization On Computer simulations \citep{CROC} to reproduce the observed modulation of $T(r)$, and concluded that the excess transmission at intermediate distances observed is not found in the CROC suite. 
Although the reason for this mismatch appears to be a too homogeneous radiation field in the simulated IGM, its root cause could not be firmly identified because the multiple explanations proposed cannot be extensively tested using expensive radiation-hydrodynamic simulations. Among the proposed sources of the mismatch in the simulations, we mention a too low mass for the smallest galaxies contributing significantly to the ionizing photons budget, an earlier reionization history, and a large stellar escape fraction. 
Additionally, the stellar-to-halo-mass relation in CROC is very flat at halo masses $M_\mathrm{halo} \gtrsim 10^{10} \, \Msun$, in disagreement with observations and theoretical models \citep[see figure 9 in \paperI and][]{Zhu+2020}, suggesting some issues with the simulated galaxy population.

We begin providing a synthetic version of the observations of \citet{Kakiichi+2018}, \citet{Meyer+2019} and \citet{Meyer+2020} in Fig.~\ref{fig:los_galaxies}, where we show the transmitted (normalized) flux at $z=5.5$ in a random sightline through the \thesanone box (solid black line and left vertical scale). 
Galaxies around the latter are shown whenever the galaxy stellar mass exceeds $M_\mathrm{star} \geq 10^{9.5} \, \Msun$ using star symbol coloured to reflect $M_\mathrm{star}$ and vertically positioned to show their distance from the sightline (right vertical scale). 
Finally, we show mock \jwst images for the ten galaxies in the plot with the largest $M_\mathrm{star}$. Notice that these are \textit{not necessarily} the largest galaxies in the simulation. 
The \jwst mock images are generated using the code \textsc{skirt} \citep[last described in ][]{skirt} and following the procedure described in \citet{Vogelsberger2020}. In particular, we show a composite image obtained combining the F277W, F356W and F444W NIRCam wide filters covering an area of $(10 \, \mathrm{pkpc})^2$. 

Moving to a quantitative analysis, we show in Fig.~\ref{fig:correlation-th1} the excess IGM transmissivity with respect to its average across all $r$, \ie $\langle T(r) \rangle / \bar{T} -1$, at $z=5.5$ in \thesanone. The comparison with observations is rendered difficult by the diverse observational techniques required to collect this type of data. 
For instance, \citet[][yellow squares in the figure]{Meyer+2019} employed the \CIV absorption along the same QSO lines of sight to estimate the position of galaxies nearby, while \citet[][cyan triangles in the figure]{Meyer+2020} employed LAE and Lyman-break galaxies.
Therefore, in an effort to enable a general comparison, we show in the figure the results obtained from \thesanone applying a number of different criteria in the galaxy identification. In particular, in the top panel of the Figure we report the curves corresponding to three minimum halo masses (black to grey curves), while in the middle panel we explore the effect of a minimum galaxy stellar mass (purple curves). In the bottom panel, we present the outcome of forward modelling the \CIV absorption features (employing the same approach used for \Lya and described in Section~\ref{sec:methods}) and using them to estimate the galaxy position along the line of sight. The orange-to-red curves report the obtained $T(r)$ for different values of the minimum \CIV column density employed in the galaxy identification, namely $\log( N_\mathrm{CIV} / \mathrm{cm}^{-2}) \geq 12.5, 13, 13.5, 14$. The first value corresponds to the nominal lower threshold imposed in \citet{Meyer+2019}, while the last corresponds to the $N_\mathrm{CIV}$ for which 90\% completeness is reached.

From the top panel of Fig.~\ref{fig:correlation-th1}, the effect of selecting galaxies hosted in haloes of different mass can be seen. 
In particular, we can partially match the observed flux modulation from \citet{Meyer+2019} only when selecting haloes with mass $M_\mathrm{halo} \gtrsim 10^{12} \, \Msun$, although the signal becomes very noisy because of the small number of haloes with such mass at $z=5.5$ in our simulation box. 
A better agreement is found when we instead select only galaxies with stellar mass $M_\mathrm{star} \geq 10^{10} \, \mathrm{M}_\odot$, as seen in the middle panel, although the recovered $\langle T(r) \rangle$ from simulations is still lower than the data from \citet{Meyer+2019}.
Consistently with the top panel, lowering the threshold in the stellar mass of selected galaxies (which effectively corresponds to lowering their host halo mass) reduces the excess flux at intermediate scales. 
This result highlights the importance of a solid characterisation of the galaxies employed to compute this correlation in order to faithfully interpret the results. While such characterisation can be achieved when surveys are employed, identifying nearby galaxies from absorption features in the same spectrum renders the task significantly more challenging. 
Finally, when employing the synthetic \CIV absorption to locate galaxies along the line-of-sight (bottom panel of Fig.~\ref{fig:correlation-th1}), we fail to reproduce the observed transmissivity modulation for all the column density thresholds employed (note that in \thesanone at $z=5.5$ there are no absorption features with $\log( N_\mathrm{CIV} / \mathrm{cm}^{-2}) \geq 14.5$). 
Before investigating the reason for this failure in the following, we note here that the \ion{C}{III} ionization frequency is just slightly higher than the ionization frequency of \ion{He}{I}. Hence, the suppression of the radiation field at such frequencies due to helium (double) reionization is highly dependent on both the details of the latter in the simulations. 
In particular, the radiation field intensity at the \CIV ionization frequency in the UVBG employed \citep[\ie][see Sec.~\ref{sec:methods}]{Faucher-Giguere+09} varies by $40$ orders of magnitude in the redshift range $6 \leq z \le 10$ (as a consequence of the evolving absorption due to helium secondary ionization. Therefore, even a small difference in the reionization history assumed in the UVBG can have catastrophic consequences for the simulated abundance of \CIV. 
We plan to quantify this effect better, and improve upon our current method, in a future work, where we will also investigate the feasibility of using the observed \CIV line evolution to constrain \highz helium reionization (and, consequently, the coeval UVBG and sources of energetic photons). 

In order to investigate the reasons for this failure, we start by showing in Fig.~\ref{fig:correlation-evol} the redshift evolution (as indicated by the line colour) of $T(r)$. For the sake of clarity, we only show two selection criteria, namely $M_\mathrm{halo} \geq 10^{10} \, \mathrm{M}_\odot$ and $M_\mathrm{star} \geq 10^{9} \, \mathrm{M}_\odot$. This choice is made to ensure that a sufficient number of galaxies is selected at each redshift. 
The curves in both panels show the same evolution, where two main effects can be appreciated. On the one hand, the signal appears stronger at higher redshift, reflecting how the average transmissivity $\bar{T}$ decreases at earlier epochs (see \eg Fig.~\ref{fig:mean_flux} for the actual mean flux evolution) much more rapidly than the transmission in the first highly-ionized regions producing the transmission spikes. 
On the other hand, the location of the excess in the transmission moves closer to the galaxies with increasing redshift. This results from the combination of two separate factors: (i) the lower overdensities in which galaxies reside at earlier times limits the suppression of the \Lya flux (because of enhanced hydrogen recombination) at smaller and smaller distances from the galaxy itself; and (ii) the smaller sizes of ionized bubbles (see \eg Fig.~\ref{fig:DG_hp_wp} or fig. 19 in \paperI), which reduce the distance at which the flux is enhanced. 
Accepting at face value the minimum \CIV host mass quoted in \citet[][\ie $M_\mathrm{halo} \geq 10^{10} \, \mathrm{M}_\odot$, corresponding to the top panel of Fig.~\ref{fig:correlation-evol}, but see their figure 2 and related discussion about the effective halo mass traced]{Meyer+2019}, we find that the match between their data and the IGM properties in \thesan is improved if we employ simulation outputs at $z\sim6.2$ in place of their mid redshift of \mbox{$\sim5.5$}, potentially indicating the necessity for an even later reionization in our model. 
However, a number of other explanations are possible \citep[see Section~\ref{sec:intro} and][for a discussion]{Garaldi+2019croc}. We next investigate the effects of the nature of dark matter and of the escape of photons from galaxies using the \thesan medium resolution runs.

\begin{figure}
\includegraphics[width=\columnwidth]{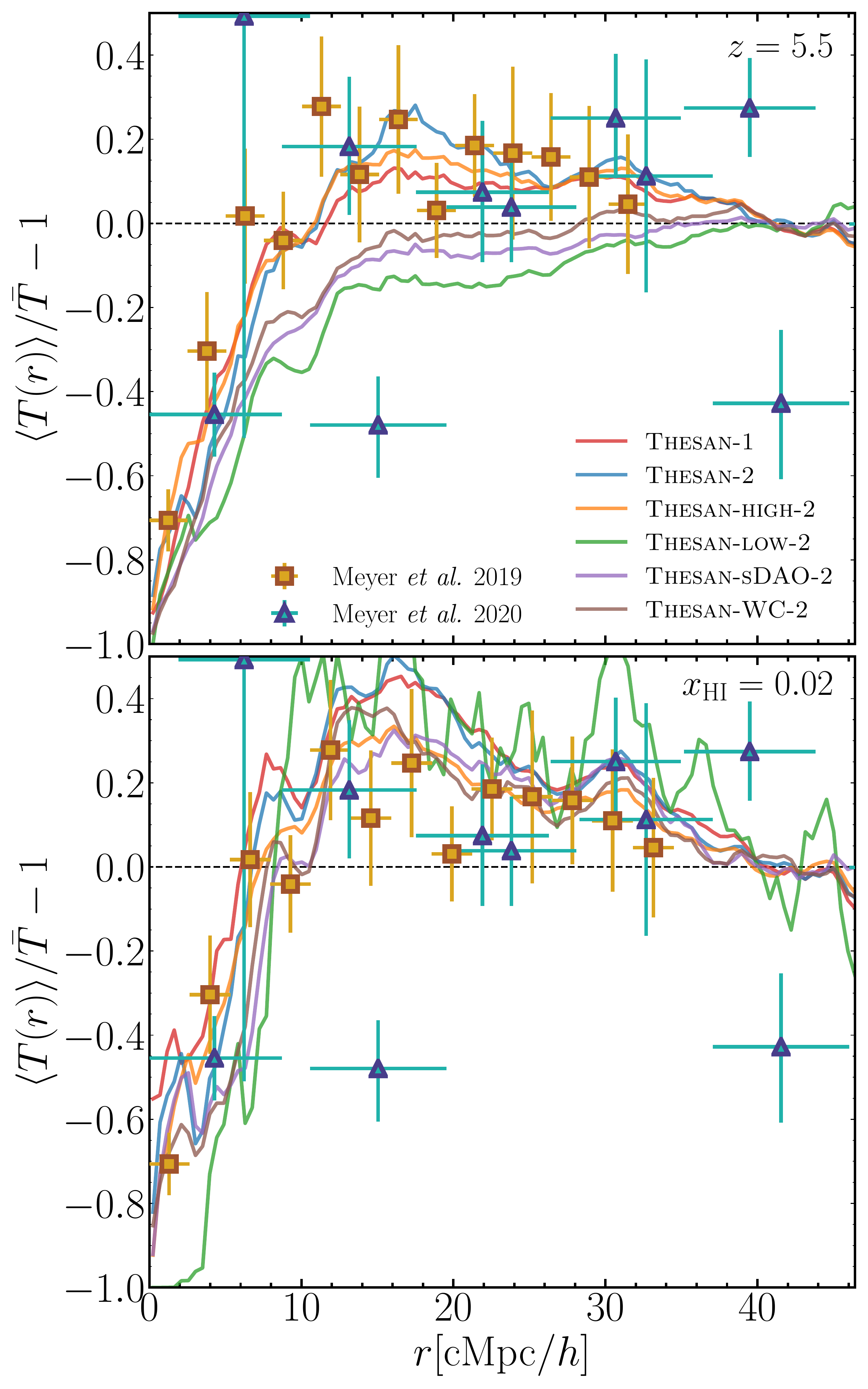}
\caption{As Fig.~\ref{fig:correlation-th1}, but now showing the impact of \highz physics at fixed redshift ($z=5.5$, top panel) and volume-averaged hydrogen neutral fraction ($x_\mathrm{HI} = 0.02$, bottom). }
\label{fig:correlation-comparison}
\end{figure}

In Fig.~\ref{fig:correlation-comparison} we show the results obtained in a selection of \thesan runs. We choose to show the curves computed selecting only galaxies with $M_\mathrm{star} \geq 10^{10} \, \Msun$, as this is the one that best matches the observed data at $z=5.5$ in \thesanone. We show the results from different runs at the same redshift (top panel) and neutral hydrogen fraction (bottom panel). 
It is immediately clear that, when a single redshift is selected, the different runs exhibit two very different types of behaviours. On one hand \thesanone, \thesantwo and \thesanhigh agree fairly well with the observed modulation, while \thesanlow, \thesansdao and \thesanwc do not exhibit any enhancement of transmission at intermediate scales, scoring very poorly in their comparison with data. 
Once we match the ionization state of the IGM, however, all curves appear very similar. While this is easily understandable for most of the models, it may be surprising that the same occurs for \thesanlow, where one could expect the signal to be reduced since only large galaxies are selected. Nevertheless, the clustering of small objects around larger ones produces a large-scale reionization topology which is similar to the one in the fiducial model in the vicinity of massive objects. This can be visually appreciated in Fig.~\ref{fig:z_reion}, where the majority of differences between \thesantwo and \thesanlow are found around small, isolated galaxies. 
Therefore, $T(r)$ appears to be robust against the physical model differences we have explored in \thesan, making it a promising tool for constraining the timing of reionization. Comparing the differences between \thesanone and \thesanwc in the top panel of Fig.~\ref{fig:correlation-comparison} to their small difference in reionization histories (see Fig.~\ref{fig:igm_prop}) it can be inferred that $T(r)$ is extremely sensitive to the reionization history of the Universe, rendering it a powerful probe of the latter, but also highlighting the necessity of more detailed studies. 
Unfortunately, the nature of this probe --~at the interface between reionization and galaxy formation and at scales where radiation feedback can play a relevant role~-- renders its study significantly expensive from a computational perspective.

\section{Summary and Conclusions}
\label{sec:conclusions}
We have introduced the \thesan project, a suite of radiation-magneto-hydrodynamic simulations designed to simultaneously and self-consistently model the Epoch of Reionization and the formation of the first galaxies. 
The simulations have box sizes of $L_\mathrm{box} = 95.5 \, \mathrm{Mpc}$, resolve atomic cooling haloes, and employ a well-tested galaxy formation model (the same developed for the IllustrisTNG suite), augmented with self-consistent radiation transport, a non-equilibrium thermo-chemistry solver, and dust creation/destruction processes. The different runs explore changes in the escape of ionizing photons from galaxies and in dark matter properties.
This paper is part of a series of three introductory manuscripts (alongside with \paperI and \paperII), and focuses on a thorough analysis of the properties of the high-redshift intergalactic medium, its connection with the galaxy population, and \Lya transmission through the IGM. 
We summarise our main results in the following points:

\begin{enumerate}
    \item The \thesan simulations have realistic reionization histories, matching most of the available constraints on the evolution of the global neutral hydrogen fraction. They all follow a `late reionization' model, completing at $z\lesssim 5.5$, with the exception of \thesanlow, where only small galaxies emit ionizing photons and the IGM is completely ionized by $z\sim6.5$. The gas temperature at mean density and the optical depth of CMB photons have realistic values in all runs (see Fig.~\ref{fig:igm_prop}).
    
    \item The low-density warm gas in \thesan shows a realistic distribution of densities and temperatures thanks to the self-consistent treatment of radiation. Its distribution (Fig.~\ref{fig:phase_space}) reflects the locally-different reionization histories (see also Fig.~\ref{fig:z_reion}) of gas regions exposed to different radiation fields. Consequently, the heat injection from photo-ionization occurs at different times in the universe, and therefore gas patches at a given redshift experienced unequal cooling, producing a wide range of temperatures.
    
    \item The number of photons produced by galaxies (as measured by the ionizing emissivity, Fig.~\ref{fig:dotN_vs_z}) appears to be only slightly larger, if extrapolated to $z\leq5$, than in available observations. The predicted evolution does not show any sign of a drop, as assumed in some recent semi-numerical studies \citep[\eg][although we postpone to a future work the investigation of the effect of an evolving ionizing photon escape fraction from haloes]{Kulkarni2019, Keating+2020}. Some of these photons make their way to the ionized IGM patches, building up the photo-ionization rate (Fig.~\ref{fig:gamma}), which appears in good agreement with observed values. The main exception is \thesanlow, which significantly over-estimates the measurements at $z \geq 5.5$.
    
     \item The properties of the \Lya forest in most \thesan runs match the observed mean flux (Fig.~\ref{fig:mean_flux}) and optical depth evolution (Fig.~\ref{fig:tau_evol_individual}) very well. When we force only small galaxies to produce ionizing photons, the mean flux is significantly higher than observed (and, consequently, the effective optical depths significantly lower) as a consequence of the early reionization history of this run. 
    
     \item The IGM in \thesanone appears slightly too homogeneous, as shown by the distribution of optical depths (Fig.~\ref{fig:tau_cdf}), which is only marginally consistent with available observations. This may be evidence in favour of a reionization history completing later than in our model (\ie at $z \lesssim 5.2$), as can be seen by looking at the same observable in \thesantwo (Fig.~\ref{fig:tau_cdf_comparison}), which has such delayed reionization with respect to the \thesanone run.
    
     \item The mean free path of ionizing photons (Fig.~\ref{fig:mfp}) shows rapid evolution between $z=5$ and $z=6$ in all runs except \thesanlow. The values computed from our main run (\thesanone) align very well with recent measurements. 
    
     \item We characterise the shape of \highz transmission regions in the \Lya forest as well as the low-flux gaps between them (Fig.~\ref{fig:DG_hp_wp}). We find they evolve significantly with redshift, opening up the possibility of constraining the history of reionization using this statistic. Additionally, we find that these measurements are sensitive to the mass of the sources powering the EoR (\ie they are able to distinguish a scenario where small galaxies produce the bulk of ionizing photons from one where large ones power the reionization of hydrogen).
    
     \item We study the modulation of the transmitted \Lya flux as a function of the distance from galaxies (Fig.~\ref{fig:correlation-th1}). This is a particularly stringent test of the simulations, since it depends simultaneously on the properties of reionization and of the galaxy population. In the past even state-of-the-art simulations could not reproduce the observed flux enhancement at intermediate scales \citep{Garaldi+2019croc}. We find that we can reproduce this general observed trend, but not its amplitude, which is smaller in \thesan than in observed QSO spectra \citep{Meyer+2019}. We investigate possible reasons for this difference and find that, consistently with other results presented, a later reionization would bring the simulation into better agreement with observed values (Fig.~\ref{fig:correlation-evol}). (Note however that the latter come from only a handful of sightlines.) Finally, we find that this measure is robust against the detailed modelling of the sources of reionization (Fig.~\ref{fig:correlation-comparison}); \ie a similar flux modulation is produced for all the models investigated, once their reionization history is accounted for.
    
\end{enumerate}

In this work we have presented the properties of the \highz intergalactic medium and \Lya transmission using the new \thesan simulation suite. We find an excellent agreement between the simulated properties and available observations, despite not explicitly tuning parameters to achieve such a result. 
The \thesan simulations represent the first radiation-hydrodynamic simulations of the `late reionization' model, which was recently suggested to explain the long Gunn-Peterson trough observed. We show that this model is also needed to account for the rapid evolution of the ionizing photon mean free path and effective optical depth distributions. 
\thesan improves on the existing literature on this subject by: including a significantly-expanded range of physical processes, reaching higher resolution, and self-consistently following the propagation of radiation on-the-fly. 
Additionally, we provide a thorough analysis of the connection between the properties of reionization and the galaxy population at $z\gtrsim5.5$, which is one of the biggest challenges that theoretical modelling will face as the properties of the first galaxies become the new frontier in studies of the EoR and galaxy formation, thanks to the flood of information that will be provided by instruments like \jwst, \alma, \hera, \ska, \ccatp and \spherex. 
We speculate that an even-later reionization history would 
likely cure most of the small discrepancies with observations. 

The simulations introduced and the forward modelling framework presented here open up many exciting prospects for the investigation of the \highz Universe. A possible avenue forward, which we plan to pursue in a future work, is to employ simulations to decode observations of high-redshift metal lines in the spectra of background QSOs. These features have the potential to unveil the production and distribution of the first metals in the Universe by the first galaxies, therefore complementing the analysis presented here, which has focused on the diffuse hydrogen. Additionally, the metal content of the IGM and CGM is another interesting research path, since recent observational campaigns have shown that virtually all metals reside in the cold gas phase at $z \gtrsim 2.5$ \citep{Peroux&Howk2020}.

In conclusion, the results described in this and in the accompanying papers provide a solid base for advancing our knowledge of cosmic reionization and \highz structure formation using the \thesan simulations, which we will make public in the near future.

\section*{Acknowledgements}
We thank the anonymous referee for their constructive comments, that improved the quality of the paper, and Benedetta Ciardi for useful comments and discussions. AS acknowledges support for Program number \textit{HST}-HF2-51421.001-A provided by NASA through a grant from the Space Telescope Science Institute, which is operated by the Association of Universities for Research in Astronomy, incorporated, under NASA contract NAS5-26555. MV acknowledges support through NASA ATP grants 16-ATP16-0167, 19-ATP19-0019, 19-ATP19-0020, 19-ATP19-0167, and NSF grants AST-1814053, AST-1814259,  AST-1909831 and AST-2007355. The authors gratefully acknowledge the Gauss Centre for Supercomputing e.V. (\url{www.gauss-centre.eu}) for funding this project by providing computing time on the GCS Supercomputer SuperMUC-NG at Leibniz Supercomputing Centre (\url{www.lrz.de}). Additional computing resources were provided by the Extreme Scisence and Engineering Discovery Environment (XSEDE), at Stampede2 through allocation TG-AST200007  and by the NASA High-End Computing (HEC) Program through the NASA Advanced Supercomputing (NAS) Division at Ames Research Center.
We are thankful to the community developing and maintaining software packages extensively used in our work, namely:  \texttt{matplotlib} \citep{matplotlib}, \texttt{numpy} \citep{numpy}, \texttt{scipy} \citep{scipy} and \texttt{cmasher} \citep{cmasher}.

\section*{Data Availability}
All simulation data, including snapshots, group and subhalo catalogues, merger trees, and high time cadence Cartesian outputs will be made publicly available in the near future. Data will be distributed via \url{www.thesan-project.com}. Before the public data release, data underlying this article will be shared on reasonable request to the corresponding author(s).

\bibliographystyle{mnras}
\bibliography{bibliography}

\appendix
\section{Comparison with the IllustrisTNG model}
\label{sec:tng_comparison}

\begin{table*}
	\centering
	\caption{Same as Table~\ref{table:simulations}, but showing the two additional runs used in this Appendix.}
	\label{table:simulations_extra}
	\addtolength{\tabcolsep}{-0.5pt}
	\begin{tabular}{lccccccccl} 
		\hline
		Name & ${L}_\mathrm{box}$ & $N_\mathrm{particles}$ & ${m}_\mathrm{DM}$ & $m_\mathrm{gas}$ & $\epsilon$ & $r^\mathrm{min}_\mathrm{cell}$& $z_\mathrm{end}$ & $f_\mathrm{esc}$ & Description\\  
		& [cMpc] & & [$\mathrm{M}_\odot$] & [$\mathrm{M}_\odot$] & [ckpc] & [pc] & & &\\
		\hline
		\thesansmall & $23.9$  & $2 \times 525^3$ & $3.12 \times 10^6$ & $5.82 \times 10^5$ & $2.2$ & $\sim 10$ & $5.5$ & $0.37$ & fiducial model (RMHD + TNG + dust) \\
		\thesansmalltng & $23.9$  & $2 \times 525^3$ & $3.12 \times 10^6$ & $5.82 \times 10^5$ & $2.2$ & $\sim 10$ & $5.5$ & - & MHD + TNG model (original TNG)\\
		\hline
	\end{tabular}
	\addtolength{\tabcolsep}{0.5pt}
\end{table*}

\begin{figure*}
\includegraphics[width=0.99\textwidth]{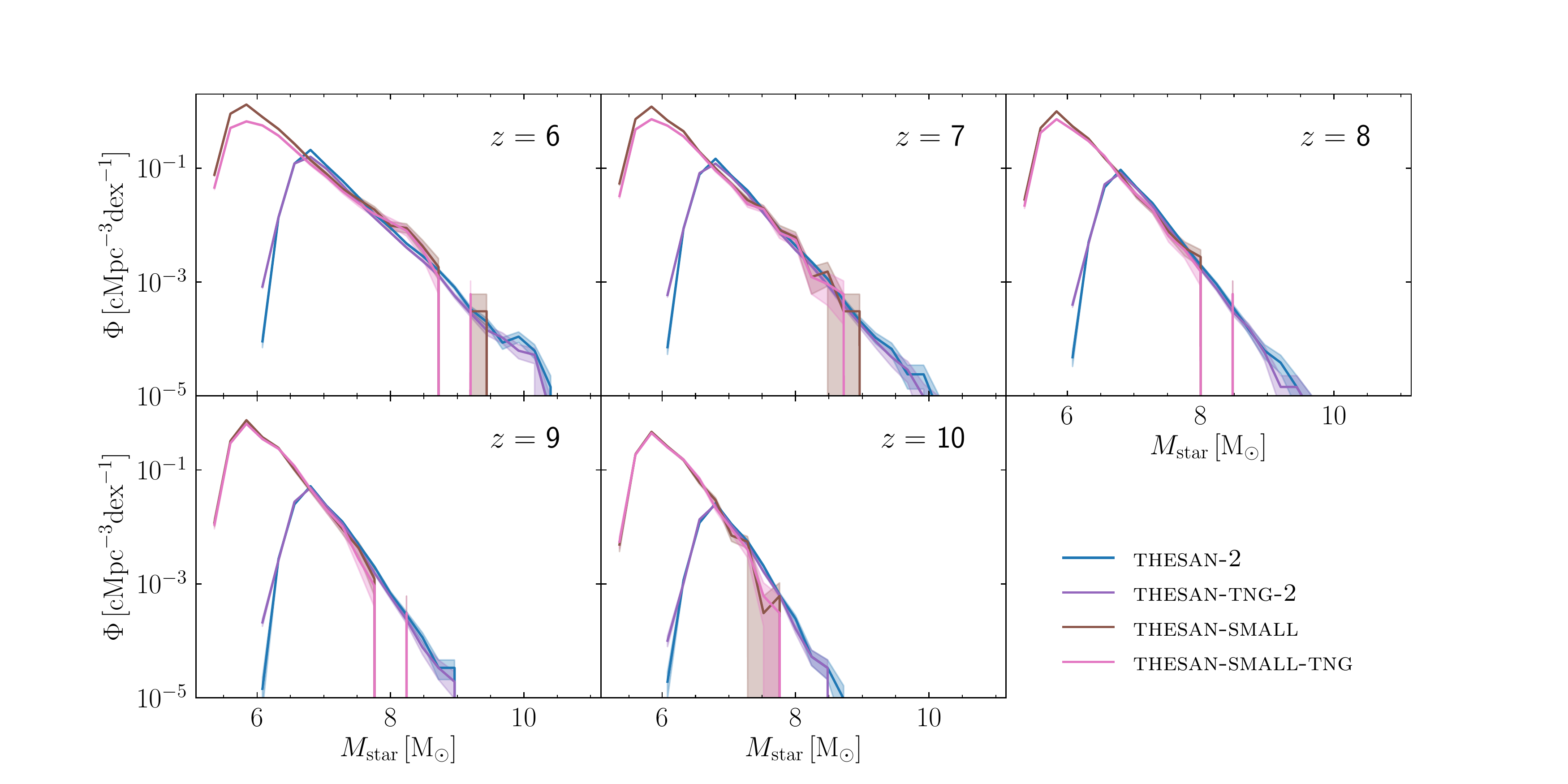}
\caption{Galaxy stellar mass function for the \thesantwo and \thesantng runs at different redshifts. The two curves differ only at low stellar masses and only towards the end of reionization.}
\label{fig:gsmf_thesan_vs_tng}
\end{figure*}

\begin{figure*}
\includegraphics[width=0.99\textwidth]{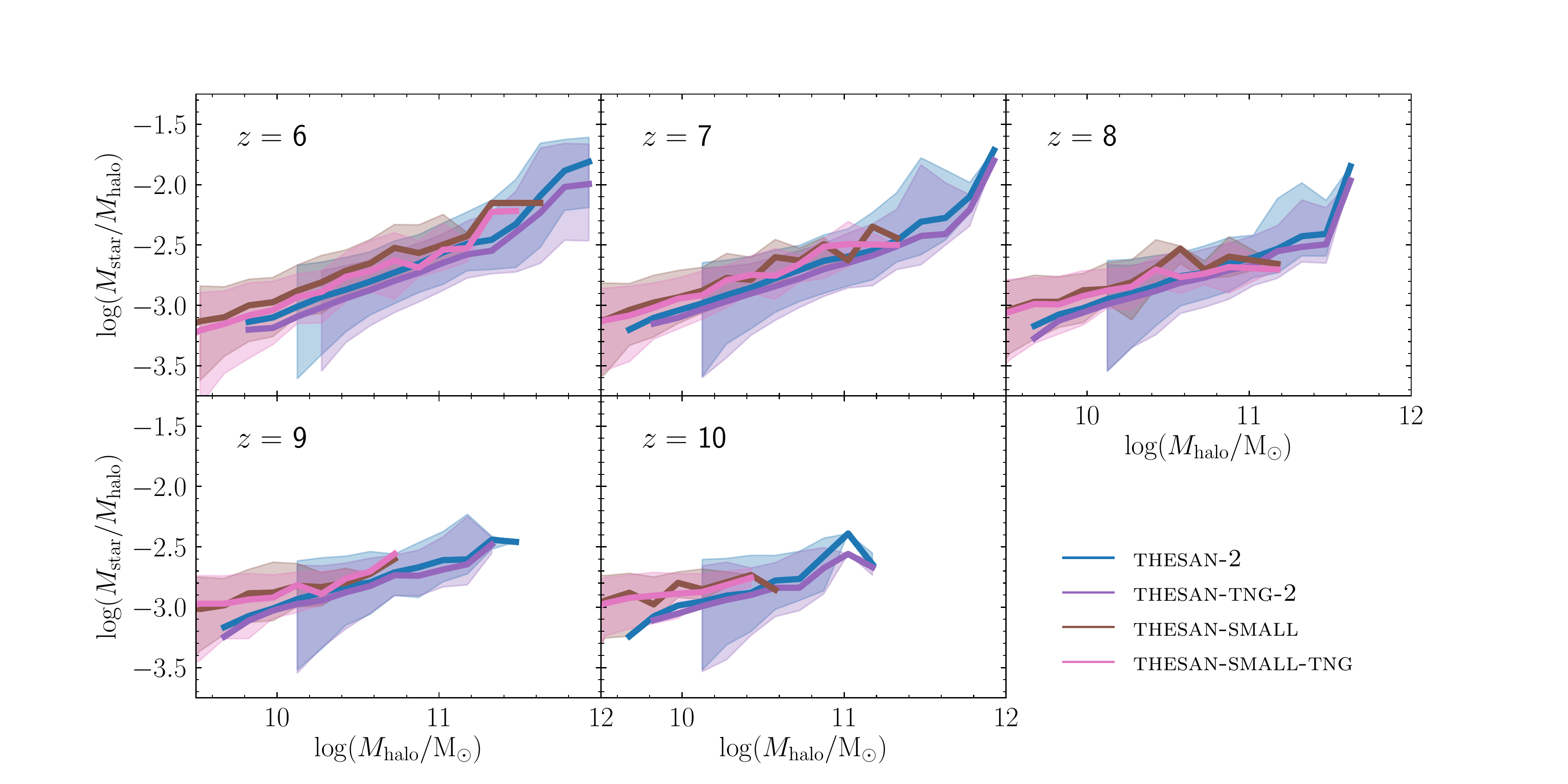}
\caption{Stellar-to-halo-mass relation for the \thesantwo and \thesantng runs at different redshifts. The two curves show a systematic offset only at $z \lesssim 8$.}
\label{fig:smhm_thesan_vs_tng}
\end{figure*}

We provide in the following a brief comparison between the \thesan model and the IllustrisTNG one, leaving a more thorough investigation to a future work. We perform such comparison by contrasting the results from the \thesantwo and \thesantng runs, that share the same initial conditions, allowing us to ascribe their differences unequivocally to the different galaxy formation models. The resolution of these runs, however, is lower than our fiducial run, potentially hiding differences in the lowest-mass haloes of \thesanone. For this reason, we have run two additional simulations covering a smaller volume but reaching the same mass resolution as our flagship run. We name \thesansmall the one employing the \thesan model, and \thesansmalltng the one using the IllustrisTNG model. They share the same initial conditions, and their properties are reported in Table~\ref{table:simulations_extra}. 

In order to confirm this interpretation, we show in Fig.~\ref{fig:smhm_thesan_vs_tng} the stellar-to-halo mass relation. This graph clearly shows that until $z=7$ the stellar content of haloes is similar in the two models, while at later times the ratio $M_\mathrm{star}/M_\mathrm{halo}$ is comparatively (slightly) enhanced in the \thesan model.

In Fig.~\ref{fig:gsmf_thesan_vs_tng} we show the galaxy stellar mass function in \thesantwo, \thesantng, \thesansmall and \thesansmalltng. 
Each pair of curves shows are very similar at all but the smallest stellar masses. At $M_\mathrm{star} \lesssim 10^6 \, \Msun$, the two models show different numbers of galaxies towards the end of reionization (\ie $z \lesssim 7$). 
In particular, the \thesan model seems to produce slightly more galaxies than the IllustrisTNG one. 
Employing a self-consistent radiation transport scheme ensures that haloes are exposed to the correct (spatially-varying) radiation field, which in the case of these small haloes is expected to be mostly generated from external sources. On the contrary, the spatially-uniform optically-thin UVBG employed by IllustrisTNG injects the same radiation in all haloes. 
Additionally, the latter is developed to complete reionization by $z\sim6$, while in \thesan this process completes later. Hence, at any given redshift, the average UVBG in \thesantwo and \thesansmall is lower than in \thesantng and \thesansmalltng, reducing the suppression of star formation in small haloes in the former. 

\begin{figure}
\includegraphics[width=\columnwidth]{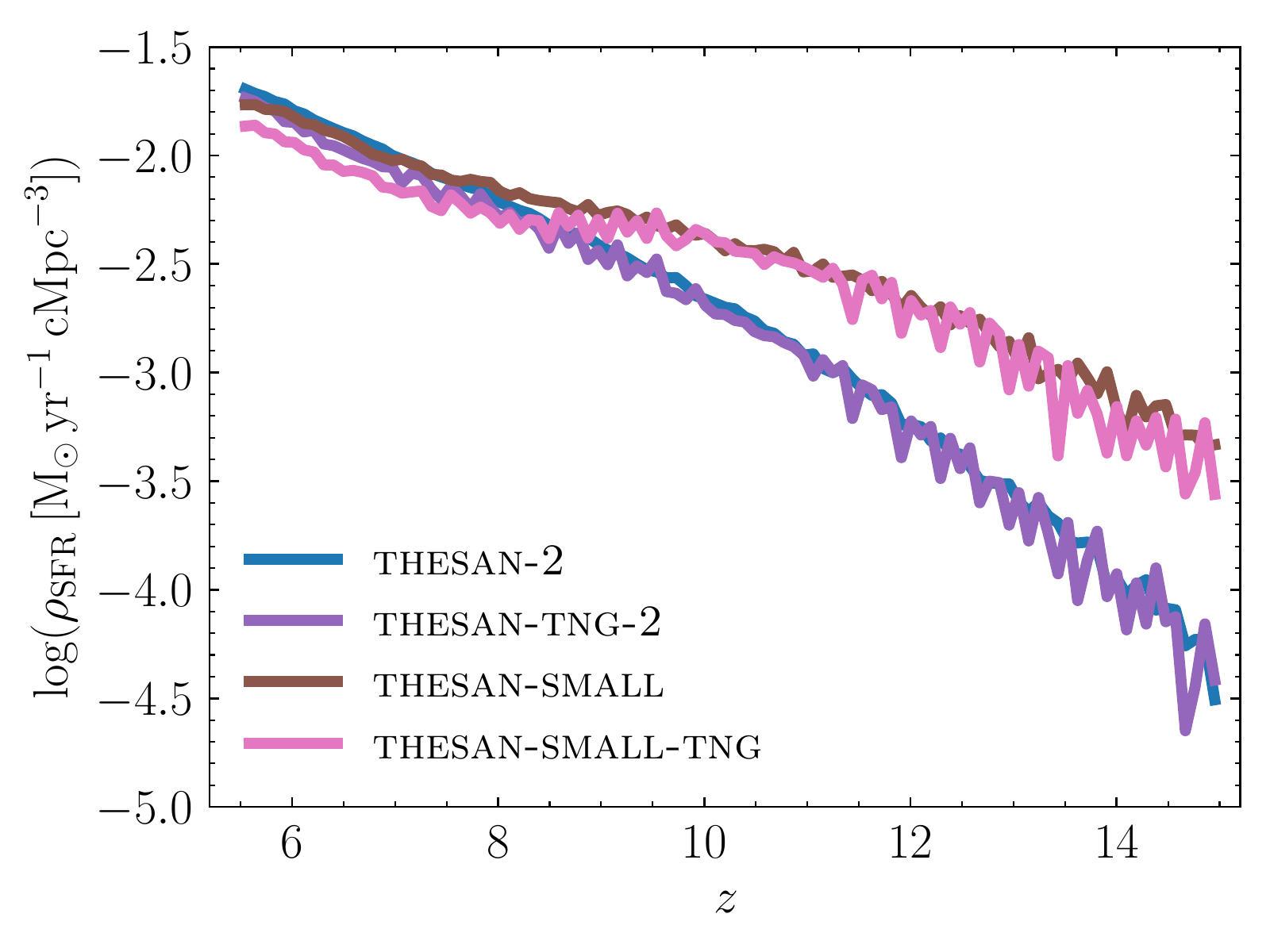}
\caption{Star formation rate density evolution for the \thesantwo and \thesantng runs. The two curves differ only at $z \lesssim 8$.}
\label{fig:sfh_thesan_vs_tng}
\end{figure}

Finally, we show in Fig.~\ref{fig:sfh_thesan_vs_tng} the evolution of the star-formation-rate density $\rho_\mathrm{SFR}$, which is indistinguishable between each pair of runs until $z\sim8$, and subsequently slightly (comparatively) suppressed in the IllustrisTNG model. The difference across the two run pairs is due to their difference resolution, which allows the \thesansmall and \thesansmalltng simulations to resolve smaller haloes. These haloes are the dominant source of star formation at very high redshift, explaining the boosted $\rho_\mathrm{SFR}$ at $z\gtrsim9$. 

\bsp 
\label{lastpage}
\end{document}